\documentclass[fleqn,twoside,twocolumn,nofootinbib]{revtex4} % Specifies the document class %,unsortedaddress
\usepackage[sec,web]{ujp}
\usepackage{xcolor}

\begin{document}
\title[Equation of State of  Quantum Gases Beyond the Van der Waals Approximation]
{Equation of State of  Quantum Gases Beyond the Van der Waals Approximation}%
\author{K.A. Bugaev}%1 автор
\affiliation{\bitp}%и
\address{\bitpaddr}%адрес
\email{bugaev@th.physik.uni-frankfurt.de}%e-mail
\email{Gennady.Zinovjev@cern.ch}
\author{A.I. Ivanytskyi}
\affiliation{Department of Fundamental Physics, University of Salamanca}
\address{Plaza de la Merced s/n 37008, Spain}
\email{oivanytskyi@usal.es}
\affiliation{\bitp}
\address{\bitpaddr}
\author{V.V.~Sagun}
\affiliation{\bitp}
\affiliation{CENTRA, Instituto Superior T$\acute{e}$cnico, Universidade de Lisboa}
\address{Av. Rovisco Pais 1, 1049-001 Lisboa, Portugal}
\email{v_sagun@ukr.net}
\author{E.G. Nikonov}
\affiliation{Laboratory for Information Technologies, JINR}
\address{Joliot-Curie str. 6, 141980 Dubna, Russia}
\email{e.nikonov@jinr.ru}
\author{G.M.  Zinovjev}
\affiliation{\bitp}
\address{\bitpaddr}
%

%%%\avtorcol{K.A. Bugaev,  A. I. Ivanytskyi, V. V. Sagun}%%колонтитул
\udk{539.12}
\razd{\secix}

\setcounter{page}{863}%
\maketitle

%%%%%%%%%%%%%%%%%%%%%%%%%%%%%%%%%%%%%%%%%%%%%%%%%%%%

%%% Beginning        06.01.2017            Kyrill     QGas_IST_1
%%% Completion      15.02.2017            Kyrill     QGas_IST_1
%%% Figures             15.04.2017             Aleksei
%%% Revisions         22-28.10.2017            Kyrill     QGas_IST.tex
%%% Revisions         Dec   2017            Kyrill     QGas_IST_newest.tex

%\title{Equation of State of  Quantum Gases Beyond the Van der Waals Approximation}

%%%%%%%%%%%%%%%%%%%%%%%%%%%%%%%%%%%%%%%%%%
%
% Do not remove these definitions below !!!!
%
%%%%%%%%%%%%%%%%%%%%%%%%%%%%%%%%%%%%%%%%% My DEFINITIONS

\def\ts{t{\ss}s}
\def\ss{\hspace{.5pt}}
%
%
%       Letters
%
% below there is an example that works for scale 1044 mode!
% small j
%\def\I2{{\large \rm \"{I}}}
%%%
%\def\si{\mbox{\footnotesize\rm \"l\hspace*{-2.25pt}l}}
%
%\def\i2{\mbox{\tiny\rm \"l\hspace*{-2.45pt}l}}

\def\i2{\mbox{\scriptsize\rm \"l\hspace*{-2.05pt}l}}

\def\bi2{\mbox{\footnotesize\rm \bf \"l\hspace*{-2.75pt}l}}

\def\ii2{\mbox{\footnotesize \it \"l\hspace*{-2.75pt}l}}

\def\I2{{\rm \"{I}}}

\newfont{\cyrfnt}{wncyr7 scaled 1400}

\newfont{\cyrftit}{wncyr8 scaled 985}
%\newcommand{\cyrb}{\baselineskip10.5pt\cyrfnt\cyracc}

%     normal   wncyr5, wncyr6, wncyr7, wncyr8, wncyr9, wncyr10
%     bold     wncyb5 .. wncyb10
%     italic   wncyi5 .. wncyi10
%     caps     wncyc5 .. wncyc10

\begin{abstract}
A recently suggested equation of state with the induced surface tension is generalized to the case of quantum gases with mean-field interaction. The self-consistency  conditions of such a model and the necessary one to obey the Third Law of thermodynamics are found.  The  quantum virial expansion of the Van der Waals models of such a  type is analyzed and its virial coefficients are given. In contrast to traditional beliefs, it is shown  that an inclusion of  the third and higher virial coefficients of the gas of hard spheres into the interaction pressure of  the Van der Waals models  either  breaks  down  the Third Law of thermodynamics or does not allow one to go beyond the Van der  Waals approximation at low temperatures.
It is demonstrated that the generalized
equation of state with the induced surface tension allows one to avoid such problems and to safely go beyond the Van der  Waals approximation.
Besides, the effective virial expansion for  quantum version of  the induced surface tension equation of state
is established and all corresponding virial coefficients are found exactly.
The explicit expressions for the true quantum virial coefficients of an arbitrary order of
this equation of state are given in the low density approximation.
A few basic constraints on such models which are necessary  to describe the nuclear and hadronic matter properties  are discussed.

\vspace*{0.5cm}

%%PACS: \, 05.70.Ce,
%%(thermodynamic functions and equations of state)
 %%21.65.mn,
%(equation of state of nuclear matter)
%%64.30.-t
%%(equations of state of specific substances)
\end{abstract}

\setlength{\topmargin}{-1.0cm}

%\maketitle

\section{Introduction} \label{Intro}
Investigation  of  equation of state (EoS) of strongly interacting particles at low temperatures is important for
studies of the nuclear liquid-gas phase transition and for properties of neutron stars \cite{GenRevEoS,Typel10,David15}.
To have a realistic EoS one has  to simultaneously  account for a short range repulsive interaction, a medium range attraction and
the quantum properties of particles. Unfortunately, it is not much known about the in-medium quantum distribution functions of  particles which experience a strong interaction.  Therefore, a working compromise to account for  all these features  is  to introduce the  quasi-particles with quantum properties which interact via the mean-field.
One of the first successful models of such  a type was a Walecka model \cite{Walecka74}.    However,  the strong demands to  consider more realistic interaction which is not restricted to some kind of effective Lagrangian led to formulating a few phenomenological generalizations
of relativistic mean-field model \cite{Zimany88,Bugaev89,Rischke91}.
Although a true breakthrough among them  was made in work  \cite{Rischke91} in which the hard-core repulsion was suggested for fermions, an introduction of phenomenological attraction in the spirit of  the Skyrme-Hartree-Fock approach \cite{SkyrmeI}  which depends not on the scalar field, but on the baryonic charge density, was also important,
since such a dependence  of attractive mean-field is typical for the EoS of  real gases  \cite{SimpleLiquids1}.

However, in addition to  the  usual defect of  the relativistic  mean-field models,  that they break down the first and the second Van Hove axioms of statistical mechanics  \cite{VanHoveI,VanHoveII},  the  usage of non-native variable, namely a particle number density, in the grand canonical ensemble  led to formulation of the self-consistency conditions \cite{Zimany88,Bugaev89}.
In contrast to the  Walecka model \cite{Walecka74} and its followers  for which  the structure of Lagrangian
and the extremum condition of the system pressure with respect to each mean-field   automatically
provide the fulfillment of the thermodynamic identities, the phenomenological mean-field EoS of hadronic matter  had to be supplemented by the self-consistency conditions \cite{Zimany88,Bugaev89,Rischke93}. The latter   allows  one to, formally,  recover the  first axiom  of statistical mechanics  \cite{VanHoveI,VanHoveII}  (for more recent discussion  of  the self-consistency conditions see \cite{Vovch14,Vovch15,Redlich16}).  An  exception  is  given by  the Van der Waals (VdW) hard-core repulsion \cite{Rischke91}, since in the grand canonical ensemble  such an  interaction  depends on the system
pressure which is the native variable for  it.

Due to its simplicity the  VdW repulsion is very popular in various branches of modern physics, but  even in  case of Boltzmann statistics it is valid only  at low particle densities for which an inclusion  of the second virial coefficient is sufficient.
For the classical gases the realistic EoS which are able to account for several virial coefficients are well-known \cite{CSEoS,SimpleLiquids1},
while a complete quantum mechanical treatment  of the third and higher  virial coefficients is rather hard \cite{Huang}.
Hence, the quantum EoS  with realistic interaction allowing one to go beyond the second virial coefficient are
of great interest not only for the dense hadronic and nuclear/neutron systems, but also for  quantum and classical liquids.
It is widely believed that one possible way to go beyond the VdW approximation, i.e. beyond the  second virial coefficient,    is to include a sophisticated interaction known from the classical models
\cite{CSEoS,SimpleLiquids1}  into the relativistic  mean-field models with the quantum distribution  functions for  quasi-particles \cite{Vovch14,Vovch15}.

{On the other hand,  a great success in getting  a high quality description of  experimental hadronic multiplicities measured in the central nuclear collisions from AGS (BNL) to LHC (CERN)
 energies  is achieved  recently with
 the hadron resonance gas model  which  employs both the traditional VdW repulsion
 \cite{Andronic06,Horn,KABOliinychenko:12,Bugaev13,Stachel:2013zma,Veta14} and  the induced surface tension  (IST) concept for the hard-core repulsion \cite{Bugaev:ALICE,VetaNEW,Bugaev17b}  {motivates} us to formulate and throughly inspect  the quantum version of this novel class of {the} IST EoS in order to apply it in the future to the  description of the properties of  dense  hadronic, nuclear, neutron matter and  dense quantum liquids on the same footing.  This is a natural choice, since the Boltzmann version of {the} IST EoS \cite{Bugaev:ALICE,VetaNEW} for a single sort of particles simultaneously   accounts for the second, third and fourth virial coefficients  of the classical gas of hard spheres and, thus, it allows one to go beyond the VdW approximation, { whereas the multicomponent formulation of such {an} EoS applied to  the mixture of nuclear fragments of all possible sizes \cite{Bugaev:13NPA} not only allows one to  introduce a compressibility of  atomic nuclei  into an exactly solvable version \cite{Bugaev:SMMI} of the statistical nuclear multifragmentation model \cite{SMM0}, but also it  sheds  light on the reason of why this model employing the proper volume approximation for the hard-core repulsion is able to correctly reproduce the low density virial expansion
 for  all atomic nuclei.}

Therefore,  the present work has two aims. First,  we would like to analyze the popular  quantum VdW models \cite{Vovch14,Vovch15,Redlich16} at high and low temperatures in order to verify whether a tuning of  interaction  allows one to go beyond the VdW treatment. In addition,  we calculate all virial coefficients for the  pressure of  point-like particles of the  quantum VdW EoS.   Second, we generalize  the recently suggested IST  EoS  \cite{Bugaev:ALICE,VetaNEW} to the quantum case,  obtain
its  effective  virial expansion and calculate all quantum virial coefficients, including the true virial coefficients for the low density limit.
Using these results,  we discuss a few basic constraints on the quantum EoS  which are necessary to model the properties of nuclear/neutron  and hadronic matter.

The work is organized as follows. In Sect. 2 we analyze  the quantum VdW EoS and its virial expansion, and   discuss the pitfalls of this EoS.  The quantum version of the IST EoS is suggested and analyzed in Sect. 3. In Sect. 4 we obtain several virial expansions of this model and discuss the Third Law of thermodynamics for the IST EoS. Some simplest applications to nuclear and hadronic matter EoS are discussed in Sect. 5,
while our conclusions are formulated in Sect 6.
}

\section{Quantum Virial Expansion for the VdW Quasi-particles} \label{VDW}

{
Similarly to the ordinary gases,  in the hadronic or nuclear systems  the source of hard-core repulsion
is connected to the Pauli blocking effect between the interacting fermionic constituents existing interior the composite particles (see, for instance,  \cite{Typel10}).
This effect appears due to the requirement of antisymmetrization of the wave function of all fermionic constituents
existing in the system and at very high densities it may lead to the Mott effect, i.e.  to a dissociation of composite
particles or even the clusters of particles  into their constituents  \cite{Typel10}.  Therefore, it is evident that at sufficiently high
densities one cannot ignore the hard-core repulsion or the finite (effective) size of composite particles and the success of
traditional EoS used in the theory of  real gases \cite{SimpleLiquids1} based on the hard-core repulsion approach
tells us that this is a fruitful framework also for quantum systems.
Hence we start from the simplest case, i.e. the quantum  VdW EoS \cite{Vovch15,Redlich16}}.
The typical form  of EoS for quantum quasi-particles of mass $m_p$ and degeneracy factor $d_p$ is  as follows
\begin{eqnarray}\label{EqI}
&&p ( T, \mu, n_{id})  = p_{id} (T, \nu (\mu, n_{id})) - P_{int} (T,  n_{id}) \, , \quad\\
\label{EqII}
&&p_{id} (T, \nu) = d_p  \int\frac{d{\bf k}} {(2\pi^3)} \frac{k^2} {3\, E (k)} \frac{1} { e^{\left( \frac{E (k)  - \nu} {T} \right)} + \zeta} \, , \\
\label{EqIII}
&& \nu (\mu, n_{id}) = \mu - b\, p + U (T, n_{id}) \, ,
\end{eqnarray}
where the constant $b \equiv 4 V_0 = \frac{16 \pi} {3} R_p^3$ is the excluded  volume of particles with the hard-core radius $R_p$ (here  $V_0$ is their  proper volume),  the relativistic energy of particle with momentum $\vec k$  is  $E (k) \equiv  \sqrt{\vec k^2+{m_p}^2}$ and the density of point-like particles is defined as $ n_{id} (T, \nu) \equiv \frac{\partial p_{id}(T, \nu)} {\partial ~ \nu}$. The parameter  $\zeta$ switches between the Fermi ($\zeta=1$), the Bose  ($\zeta=-1$) and the Boltzmann  ($\zeta=0$)
 statistics.  The interaction part of pressure $P_{int} (T,  n_{id}) $ and the  mean-field $U (T, n_{id})$ will be specified later.

 {  Note that similarly to the Skyrme-like EoS and the EoS of real gases    it is assumed that  the  interaction between quasi-particles
 described by the system (\ref{EqI})-(\ref{EqIII})  is completely accounted by the excluded volume (hard-core repulsion), by  the  mean-field  potential  $U (T, n_{id})$ and by the  pressure $P_{int} (T,  n_{id}) $.  This is  in contrast to the
 relativistic mean-field models of Walecka type in which   the mass shift of quasi-particles is taken into account.
Since such an effect may be important for the modeling the chiral symmetry restoration in hadronic matter the strongest arguments of whose
existence are  recently given  in  \cite{Bugaev17b},
we leave it   for a future exploration and concentrate here on a simpler  EoS defined by Eqs. (\ref{EqI})-(\ref{EqIII}).

 The functions  $U (T, n_{id})$ and  $P_{int} (T,  n_{id}) $ are not independent,  due to
 the thermodynamic identity $ n (T, \nu (\mu, n_{id})) \equiv \frac{\partial p (T, \nu (\mu, n_{id}))} {\partial  \mu}$.  Therefore,
  the mean-field terms $U$ and $P_{int}$ should obey the self-consistency condition
\begin{eqnarray}\label{EqIV}
&&n_{id} \frac{\partial U (T, n_{id})} {\partial  n_{id}} =  \frac{\partial P_{int} (T, n_{id})} {\partial  n_{id}}
  \quad  \Rightarrow  \\
&&P_{int} (T, n_{id})  = n_{id} \, U (T, n_{id}) - \int\limits_0^{n_{id}} d n \,  U (T, n)
\,, \quad  \quad
\label{EqV}
\end{eqnarray}
After integrating by parts Eq. (\ref{EqIV}),  we used in (\ref{EqV}) an obvious condition $U (T, 0) < \infty$.  If the condition (\ref{EqV}) is obeyed,  then the direct calculation of the $\mu$-derivative  of the pressure (\ref{EqI})  gives one the usual
expression for particle number density  in terms of the density of point-like particles
\begin{eqnarray}
\label{EqVI}
&&n =  \frac{n_{id} } {1 + b\, n_{id}} \,, \\
&&n_{id} (T, \nu) =  d_p  \int\frac{d{\bf k}} {(2\pi^3)}  \frac{1} { e^{\left( \frac{E (k)  - \nu} {T} \right)} + \zeta}
 \,.
 \label{EqVII}
\end{eqnarray}
From these equations one finds that $n   \rightarrow b^{-1}$ for  $n_{id} \rightarrow \infty$. The limit  $n_{id} \rightarrow \infty$ is provided by the conditions
 $\nu \rightarrow \infty$ or $T\rightarrow \infty$ for $\zeta = \{ 0;  1\}$, while for $\zeta = -1$  it is provided by the conditions   $\nu \rightarrow m_p-0$ or $T\rightarrow \infty$.

Note that in contrast to other works discussing Eqs. (\ref{EqIV}) and (\ref{EqV}) through this paper we will use the density of point-like particles $n_{id}$ as an argument of the  functions $U (T, n_{id})$ and  $P_{int} (T,  n_{id}) $ instead of the physical density of particles  $n$ because for more sophisticated EoS their relation will be more complicated than  (\ref{EqVI}).   Also such a representation is convenient for a  subsequent analysis  because in terms
of  $n_{id}(T, \nu)$ the virial expansion of $p_{id} (T, \nu)$ looks extremely simple \cite{Huang}
\begin{eqnarray}\label{EqVIII}
&&p_{id} (T, \nu ) =  T \, \sum\limits_{l=1}^\infty  a_l^{(0)} \left[ n_{id} (T, \nu )  \right]^l \,,
  \quad  {\rm where}  \\
  \label{EqIX}
&& a_1^{(0)} =1\,,\\
\label{EqX}
&&a_2^{(0)} = -  b_2^{(0)}\,,\\
\label{EqXI}
&&a_3^{(0)} = 4 \left[ b_2^{(0)} \right]^2 - 2\, b_3^{(0)} \,,\\
\label{EqXII}
&&a_4^{(0)} = -20  \left[ b_2^{(0)} \right]^3  +  18\, b_2^{(0)}\, b_3^{(0)} - 3 \, b_4^{(0)}\,, \\
&&\dots\dots
\end{eqnarray}
Here the  first few virial coefficients $a_l^{(0)}$ of  an ideal quantum gas are expressed in terms of the corresponding cluster integrals $b_{l>1}^{(0)}$ which depend only on temperature. The latter can be expressed via the  thermal density of  the auxiliary Boltzmann system $n_{id}^{(0)}(T, \nu) \equiv  n_{id} (T, \nu)|_{\zeta =0} $  of  Eq. (\ref{EqVII}) \cite{Huang,Kostyuk00}
\begin{eqnarray}
 \label{EqXIV}
&& b_l^{(0)}  =  \frac{(\mp1)^{l+1}} {l}  {n_{id}^{(0)}(T/l, \nu) } {\left[ n_{id}^{(0)}(T, \nu)  \right]^{-l}} \,,
\end{eqnarray}
where the upper (lower) sign corresponds to Fermi (Bose) statistics. For the non-relativistic case   the expression   (\ref{EqXIV}) can be further simplified  \cite{Huang} and for an arbitrary degeneracy factor $d_p$ it acquires the
form  \cite{Kostyuk00}
\begin{eqnarray}
 \label{EqXV}
&& b_l^{(0)} \biggl|_{nonrel} \simeq \,   \frac{(\mp1)^{l+1}} {l^\frac{5} {2}} \left(\frac{1} {d_p} \left[ \frac{2\, \pi} {T \, m_p} \right]^\frac{3} {2}\right)^{l-1} \,.
\end{eqnarray}
For  high temperatures one can write an ultra-relativistic analog of Eq.  (\ref{EqXV}) for a few values of  $l =2, 3, ... \ll T/m_p$
\begin{eqnarray}
 \label{EqXVI}
&& b_l^{(0)} \biggl|_{urel} \simeq \,   \frac{(\mp1)^{l+1}} {l^4} \left[ \frac{\pi^2} {d_p\,T^3} \right]^{l-1} \,.
\end{eqnarray}
Suppose that the coefficients $a_l^{(0)}$ from Eq. (\ref{EqVIII}) are known and that the virial expansion is convergent for
the considered $T$. Then using  Eq.  (\ref{EqVI})  one finds  $n_{id} = n/(1 - b\, n)$ and, hence,  one can rewrite Eq. (\ref{EqVIII})  as
\begin{eqnarray}\label{EqXVII}
&& \frac{p_{id} (T, \nu )} {T \, n} =   \frac{1} {1 - b \, n}  +  \sum\limits_{l=2}^\infty  a_l^{(0)}\frac{ \left[ n  \right]^{l-1} } {\left[1- b \, n  \right]^l } \,.
\end{eqnarray}
{Note that the expansions of such a type for a system pressure which use the variable $n/(1 - b\, n)$ instead of $n$ are well-known for the hard discs \cite{HardD}
and  hard spheres \cite{HardS} EoS, since they provide very fast convergence of the series due to very fast decrease of their coefficients.  }

As one can see from Eqs. (\ref{EqXV}) and (\ref{EqXVI}) at high temperatures all cluster integrals and virial  coefficients
of ideal quantum gas  strongly decrease with the temperature $T$ and, hence, at high temperatures the virial expansion of
$p_{id} (T, \nu )$ is defined by the first (classical)  term on the right hand side of   (\ref{EqXVII}), i.e. in this case one gets
\begin{eqnarray}\label{EqXVIII}
\hspace*{-3mm}&& \frac{p_{id} (T, \nu )} {T \, n} \simeq  1 +  4 V_0 \, n + (4 V_0 \, n)^2 + (4 V_0 \, n)^3  + \,  ... , ~\quad~
\end{eqnarray}
where after expanding the first term on the right hand side of  (\ref{EqXVII}) we used the relation between $b$ and $V_0$.
From this equation one sees that  only the  second virial coefficient, $4 V_0$, coincides with the one for the gas of hard spheres,
while the third, $16 V_0^2$, and the fourth, $64 V_0^3$ virial coefficients are essentially larger  than their counterparts
$B_3=10 V_0^2$ and  $B_4= 18.36 V_0^3$ of the gas of hard spheres.  Also Eq. (\ref{EqXVII}) can naturally  explain why in the work  \cite{Vovch14}
the authors insisted on the interaction pressure $P_{int}$ to be a linear function of $T$ (see a statement after Eq. (62) in \cite{Vovch14}): if one chooses
the interaction pressure in the form
\begin{eqnarray}\label{EqXIX}
\hspace*{-3mm}&& P_{int} (T,  n(n_{id}) )  = T F (n (n_{id})) =  T n \times \nonumber  \\
\hspace*{-3mm}&&    \left[ (b^2-B_3) n^2 + (b^3-B_4) n^3 + (b^4-B_5) n^4 + ...\right] \,, \quad \quad
\end{eqnarray}
then at high temperatures the quantum corrections are negligible and, hence, for such a choice of  $P_{int} (T,  n(n_{id}) )$
with  the corresponding value for the mean-field potential $U(T,  n(n_{id}) )$ obeying  the self-consistency condition (\ref{EqIV}), one can improve the total pressure of mean-field model by matching its repulsive part  to  the pressure  of hard spheres.

The problem, however, arises at low temperatures, while calculating the entropy density for the model with  $P_{int} (T,  n(n_{id}) )$  (\ref{EqXIX}).  
Indeed, for any  choice of {the} mean-field potential of the form  $U(T,  n(n_{id}) )=g(T) f(n(n_{id}))$  (note that Eq. (\ref{EqXIX}) has such a form) from the thermodynamic identities $s = \frac{\partial p (T, \mu )} {\partial  T}$ and
$s_{id} = \frac{\partial p_{id} (T, \nu )} {\partial  T}$ one finds \cite{Vovch14}
\begin{eqnarray}\label{EqXX}
\hspace*{-3mm}&& s (T, \mu )   = \left[{\textstyle s_{id} + \left[n_{id} \frac{\partial U} {\partial  T} -  \frac{\partial P_{int}} {\partial  T}  \right]} \right] \left[1 + b \, n_{id}\right]^{-1} = ~\quad~ \\
\label{EqXXI}
\hspace*{-3mm}&&  \biggl[ s_{id} +  \frac{d   g(T)} {d ~ T} \int\limits_0^{n_{id}} d \tilde n \,  f (n(\tilde n)) \biggr] \left[ 1 + b \, n_{id} \right]^{-1}  \,, ~\quad~
\end{eqnarray}
where in deriving Eq. (\ref{EqXXI}) from Eq.  (\ref{EqXX})  we used  Eq.  (\ref{EqV}) to express the interaction pressure 
 $P_{int}$ in terms of the potential $U(T,  n(n_{id}) )=g(T) f(n(n_{id}))$.  Using such an expression one finds  the following derivative
 \begin{eqnarray}
\frac{\partial P_{int}} {\partial  T}  =  n_{id} f(n(n_{id}))  \frac{d g(T)} {d~ T} -  \frac{d   g(T)} {d  T} \int\limits_0^{n_{id}} \hspace*{-1.1mm} d \tilde n   f (n(\tilde n)) \,.
\end{eqnarray}
Substituting this expression into   (\ref{EqXX})  one gets  Eq. (\ref{EqXXI}).

As one can see now from Eq.   (\ref{EqXXI}) the mean-field model with the  linear $T$ dependence of  $U$ or, equivalently, of  $P_{int}$, i.e. $g(T) = T \Rightarrow  \frac{d   g(T)} {d ~ T} = 1$, breaks down the Third Law of thermodynamics, since at $T=0$ one finds $ s_{id} (T=0, \nu) = 0$ by construction,  whereas  for the full entropy density one gets
\begin{eqnarray*}
s (T=0, \mu ) =   \left[ 1 + b \, n_{id} \right]^{-1} \frac{d   g(T)} {d ~ T} \cdot \int\limits_0^{n_{id}} d \tilde n \,   f (n(\tilde n)) \neq 0,
\end{eqnarray*}
\noindent
unless  $f   \equiv 0$. Hence, the mean-field model with   the linear  $T$ dependence of  $P_{int}$ suggested in  \cite{Vovch14}    may be very good at high temperatures, for which the Boltzmann statistics is valid, but it is unphysical at $T = 0$.  

Of course, one can repair this defect by choosing  more complicated function $g(T)$, which  
at high $T$ behaves as
 $g(T) \sim T$, but its derivative $g^\prime(T)$ vanishes at $T =0$ providing the fulfillment of the Third Law of thermodynamics   (see an example in Sect. 5 for  which $g(T) \sim T^2$ at low temperatures).  However, in this case the whole idea to compensate the  defects of
the VdW EoS by tuning the interacting part of pressure  does not work at low $T$,  since   in this case  $P_{int} = g(T)  F(n_{id})$ would vanish faster than the first term  staying on the right hand side of  Eq. (\ref{EqXVII}), i.e. the classical part of the pressure $T n_{id} = T n /(1-bn ) $.
{\it Thus, we explicitly
showed here that at low $T$  the mean-field models defined by Eqs. (\ref{EqI})-(\ref{EqV})  either are unphysical, if $P_{int} = T  F(n_{id})$,  or they cannot go beyond the VdW approximation by adjusting their interaction  pressure  $P_{int}$.}

{Such a conclusion can be also applied to the one of two ways to introduce   the  excluded volume  correction into the quantum second virial
coefficients discussed in Ref.  \cite{Typel16}. Although the model of  Ref.  \cite{Typel16} contains the scalar  mean-fields which modify the masses of particles,  the effective potential  approach to treat the excluded volume  correction of  Ref.  \cite{Typel16} with the linear
$T$ dependence of the repulsive effective potential $W_i$  (equivalent to the mean-field potential $-U$ in our notations) of the $i$-th particle sort (see Eqs. (20) and (46) and (47) in  \cite{Typel16})
should unavoidably lead to a  break down of the Third Law of thermodynamics.  Therefore, we conclude that such a way
to introduce   the  excluded volume  correction into the quantum second virial
coefficients discussed in   \cite{Typel16} is unphysical. Thus, despite the claims of author   of  Ref. \cite{Typel16}  such a generalization
of  the approach \cite{Rischke91} to include the hard-core  repulsion in quantum systems
leads to a problem with the Third Law of thermodynamics. 
}

To end this section we express  the traditional virial coefficients  $a_k^{Q}$ of the  quantum VdW gas of Eq. (\ref{EqXVII})  in terms of the classical excluded volume  $b$ and the quantum virial coefficients of point-like particles $a_k^{(0)}$. Expanding each denominator in Eq. (\ref{EqXVII})  into a series of  powers of $n$,  one can easily find
\begin{eqnarray}
\label{EqXXIIa}
&& p_{id} (T, \nu ) =  T \, \left[ n + \sum\limits_{k=2}^\infty  a_k^{Q} n^k  \right] \,,
  \quad  {\rm where}  \\
\label{EqXXIIb}
&&a_2^{Q} =  b + a_2^{(0)}\,,\\
\label{EqXXIIc}
&&a_3^{Q} =  b^2 + 2\, b \, a_2^{(0)}+ a_3^{(0)} \,,\\
\label{EqXXIId}
&&a_4^{Q} =  b^3 + 3\, b^2 \, a_2^{(0)}+ 3\, b^1 \, a_3^{(0)} + a_4^{(0)} \,,\\
\label{EqXXIIe}
&&a_k^{Q} =  b^{k-1} + \sum\limits_{l=2}^{k} \frac{(k-1)!}{(l-1)! (k-l)!}  b^{k-l} a_l^{(0)}  \,.
\end{eqnarray}
If the interaction  pressure $P_{int} (T,  n_{id}(n) )$ of the model  (\ref{EqI}) can be expanded into
the Taylor series of  particle number density $n$ at $n=0$, then one can obtain the full quantum virial
expansion of  this EoS.  Note that the coefficients $a_k^{(0)}$ for the model  (\ref{EqI}) depend  on temperature only, while
specific features of the EoS are stored in $b$ and in   $P_{int} (T,  n_{id}(n) )$.
For example, using the coefficients $b=3.42$ fm$^3$ and  $P_{int} (T,  n ) = a_{attr} n^2$ ($a_{attr} =329$ MeV$\cdot$ fm$^3$)
found in \cite{Vovch15} for the quantum VdW EoS of nuclear matter, one can calculate the full quantum second virial coefficient of the model as
\begin{eqnarray}
\label{EqXXIIf}
&&
%%\hspace*{-0.9cm}
a_2^{Q, tot} = \nonumber \\
&&
%%\hspace*{-0.9cm}
b + a_2^{(0)} - \frac{a_{attr}}{T} \simeq b + \frac{1}{2^\frac{5} {2} d_p} \left[ \frac{2\, \pi} {T \, m_p} \right]^\frac{3} {2}  - \frac{a_{attr}}{T} \,,
\end{eqnarray}
where in the second step of derivation we used the non-relativistic expression for the cluster integral $b_2^{(0)}$ (\ref{EqXV}).
Taking results from  \cite{Vovch15}, one can find that for nucleons $(d_p=4, m_p= 939 {\rm \, MeV})$ the coefficient $a_2^{Q, tot} (T)$ is zero at $T\simeq 0.32$ MeV and $T\simeq 90.5$ MeV, is negative between these temperatures   and then above $T\simeq 90.5$ MeV it grows  almost linearly with $T$   to $a_2^{Q, tot} (T=150 {\rm \, MeV}) \simeq (3.42 + 0.101 - 2.19) {\rm \, fm}^3 \simeq 1.33$ fm$^3$  which corresponds to  the equivalent hard-core radius $R_{eq} \simeq 0.46$ fm at $T=150$ MeV. From this estimate  it is evident that the large value of
 the equivalent hard-core radius $R_{eq}$ for the model  \cite{Vovch15} is a consequence of the unrealistically large hard-core radius  of nucleons  $R_n \simeq 0.59$ fm obtained in  \cite{Vovch15} (also, see a discussion later), whereas in the most advanced version of the hadron resonance gas model the hard-core radius of nucleons is 0.365 fm \cite{Bugaev:ALICE,VetaNEW,Bugaev17b} and in the IST EoS of the nuclear matter this radius is below 0.4 fm \cite{Aleksei17}. It is obvious that  more realistic attraction than the one used in   \cite{Vovch15}  would decrease   the values of  $R_{eq}$ and $R_n$ to physically more adequate ones.
Although the explicit  quantum virial expansion (\ref{EqXXIIa})-(\ref{EqXXIIf}) can be used to find the appropriate attraction in order to cure the problems of the VdW EoS
 and extend it to higher particle number densities and high/low $T$ values,  the true solution of this problem is suggested below.

%%%%%%%%%%%%%%

\section{EoS with Induced Surface Tension} In order to overcome the difficulties  of  the quantum VdW EoS
at high particle number densities  we suggest the following EoS
\begin{eqnarray}
\label{EqXXII}
&& p = p_{id}(T, \nu_1  )  - P_{int\,1} (T,  n_{id\,1})  \,,\\
\label{EqXXIII}
&&\Sigma =  R_p \left[  p_{id}(T, \nu_2 )  - P_{int\,2} (T,  n_{id\,2})  \right]\, , \quad \\
\label{EqXXIV}
&&\nu_1 = \mu - V_0\,p-  S_0\, \Sigma + U_1 (T,  n_{id\,1})\,, \\
\label{EqXXV}
&& \nu_2 = \mu - V_0\,p-  \alpha S_0\, \Sigma + U_2 (T,  n_{id\,2}) \,, \quad
\end{eqnarray}
where $n_{id\,A} \equiv \frac{\partial p_{id}(T, \nu_A)} {\partial ~ \nu_A}$ with $A=\{1; 2\}$,  $S_0 = 4 \pi R_p^2$ denotes the proper surface of the  hard-core volume $V_0$. Eq.  (\ref{EqXXII})  is an  analog of Eq. (\ref{EqI}), while the equation for the induced surface tension coefficient  $\Sigma$ (\ref{EqXXIII}) was first introduced
for the Boltzmann statistics in  \cite{Bugaev:13NPA}. The  system   (\ref{EqXXII})-(\ref{EqXXV})  is a quantum  generalization  of  the Boltzmann EoS  in the spirit of work \cite{Rischke91}.
{As it was argued above the  temperature dependent effective potentials  considered  in \cite{Typel16}
may lead to an unphysical behavior  at low temperatures and, hence, below  we would like to study this problem in details.
Also below we will show what is a principal difference of the EoS  (\ref{EqXXII})-(\ref{EqXXV}) with the second way to include the hard-core repulsion in quantum systems discussed in Ref.  \cite{Typel16}.
}

The quantity $\Sigma$ defined by  (\ref{EqXXIII}) is the surface part of the hard-core repulsion \cite{VetaNEW}.
As it will be shown later,  representing  the hard-core repulsion  in pressure (\ref{EqXXII}) in two terms, namely via  $-V_0 p$ and $- S_0 \Sigma$, instead of  a single term  $-4 V_0 p$ as it is done in the quantum VdW EoS, has great advantages and allows one to go beyond the VdW approximation.

Evidently, the self-consistency conditions  for the IST EoS are similar to Eqs. (\ref{EqIV}) and  (\ref{EqV}) ($A= \{1; 2\}$)
\begin{eqnarray}
\label{EqXXVI}
&&
%%\hspace*{-5mm}
n_{id\,A} \frac{\partial U_A (T, n_{id\,A})} {\partial  ~n_{id\,A}} =  \frac{\partial P_{int\, A} (T, n_{id\,A})} {\partial ~ n_{id\,A}} \, ,
\end{eqnarray}

The model parameter $\alpha > 1$ is a switch between the excluded and proper volume  regimes. To demonstrate this property let us
consider the quantum distribution function
\begin{eqnarray}\label{EqXXVII}
\hspace*{-3mm}&&\phi_{id} (k, T, \nu_2)   \equiv    \frac{1} { e^{ \frac{E (k)  - \nu_2} {T} } + \zeta} =  \nonumber \\
\hspace*{-3mm}&&\frac{e^\frac{\nu_2-\nu_1} {T}} { e^{ \frac{E (k)  - \nu_1} {T} } +  \zeta - \zeta \left[1 - e^\frac{\nu_2-\nu_1} {T}  \right] }
= \phi_{id} (k, T, \nu_1)\, e^\frac{\nu_2-\nu_1} {T} \times \quad \nonumber \\
\hspace*{-3mm}&&\biggl\{ 1 +   \sum\limits_{l=2}^\infty  \left[\phi_{id} (k, T, \nu_1) \, \zeta \, \left(1 -   e^\frac{\nu_2-\nu_1} {T} \right)\right]^l  \biggr\} \, ,
\end{eqnarray}
where in the last step of the  derivation we have expanded the longest denominator above into a series of $ \phi_{id} (k, T, \nu_1) \, \zeta \, \left(1 -   e^\frac{\nu_2-\nu_1} {T} \right)$ powers. Consider two limits of  (\ref{EqXXVII}), namely   $e^\frac{\nu_2-\nu_1} {T} \simeq 1$ and  $e^\frac{\nu_2-\nu_1} {T} \rightarrow 0$ for $\zeta \neq 0$. Then  the distribution function (\ref{EqXXVII}) can be cast as:
\begin{eqnarray}\label{EqXXVIII}
%%\hspace*{-7mm}
&& \phi_{id} (k, T, \nu_2) \rightarrow  \nonumber \\
%%\hspace*{-7mm}
&& \phi_{id} (k, T, \nu_1)\, e^\frac{\nu_2-\nu_1} {T}
 \left\{
\begin{array} {lll}
{\rm for} ~  \zeta \neq 0 \, ,  &  {\rm if} ~e^\frac{\nu_2-\nu_1} {T} \simeq 1 \,,   \\
 %&   \\
%
{\rm for} ~  \forall ~\zeta  \, ,  &  {\rm if} ~e^\frac{\nu_2-\nu_1} {T} \rightarrow 0  \,.
\end{array}
\right.
\end{eqnarray}
Further on we assume that the inequality
\begin{eqnarray}\label{EqXXIX}
\hspace*{-3mm}&&  (\alpha-1) S_0\, \Sigma / n_{id\,2} \gg  (U_2 - U_1) / n_{id\,2} \,, ~\quad~
\end{eqnarray}
holds in either of the considered limits for
%a parameter
$e^\frac{\nu_2-\nu_1} {T}$.
Note that for  the case  $e^\frac{\nu_2-\nu_1} {T} \simeq 1$  the condition (\ref{EqXXIX}) is a natural one because
 at low particle densities it means that  the difference of two mean-field potentials  $(U_2 - U_1)$  is  weaker than the hard-core repulsion term $ (\alpha-1) S_0\,  \Sigma $; whereas for  $e^\frac{\nu_2-\nu_1} {T} \rightarrow 0$
it means that such a  difference  is  simply restricted from above for large values of $\Sigma$, i.e.  $\max\{|U_1|;  |U_2|\} < Const < \infty$. Evidently, in this limit the mean-field pressures should be also finite,
 i.e. $|P_{int\,A}| < \infty$.

For the  case  $e^\frac{\nu_2-\nu_1} {T} \simeq 1$ one immediately recovers the following relation 
\begin{eqnarray*}
p_{id}(T, \nu_2 ) \simeq  e^\frac{(1-\alpha) S_0\, \Sigma } {T}  p_{id}(T, \nu_1  )
\end{eqnarray*}
for  $\zeta \neq 0$, which exactly corresponds to the Boltzmann statistics version   \cite{VetaNEW} of the system  (\ref{EqXXII})-(\ref{EqXXV}) and, hence,   one  recovers the virial
expansion of  $p_{id}(T, \nu_1)$  \cite{VetaNEW}  in terms of  the  density  of particle number $n_{1} = \frac{\partial p_{id}(T, \nu_1)} {\partial ~\mu}|_{U_1}$
which is calculated  under the  condition $U_1 = const$
\begin{eqnarray}
\label{EqXXX}
&&\hspace*{-4mm}\frac{p_{id}(T, \nu_1)} { T n_{1}}  \simeq     1 + 4 V_0 n_{1} + \left[16 - 18 (\alpha-1) \right]\, V_0^2 \,n_{1}^2 +  \nonumber \\
&&\hspace*{-4mm}\left[64 + \frac{243} {2}
(\alpha-1)^2 - 216 (\alpha-1) \right]\, V_0^3 \, n_{1}^3 + ... \,.
\end{eqnarray}
Note that due to the self-consistency condition  (\ref{EqXXVI}) one finds
$\frac{\partial p (T, \nu_1)} {\partial ~\mu}= \frac{\partial p_{id}(T, \nu_1)} {\partial ~\mu}|_{U_1}$, and, therefore, $n_{1}$  is the physical particle number density.

As it was  revealed  in   \cite{VetaNEW} for $\alpha = \alpha_{B} \equiv 1.245$  one can reproduce the fourth virial coefficient of the gas of hard spheres  exactly, while
the value of the third virial coefficient of such a gas is recovered with the relative error about $16$\% only. Therefore, for low densities, i.e. for $V_0 n_{1} \ll 1$, the IST EoS  (\ref{EqXXII})-(\ref{EqXXV}) reproduces the results obtained for $\zeta=0$, if the condition (\ref{EqXXIX}) is fulfilled.

On the other hand, from Eqs.  (\ref{EqXXVII}) and  (\ref{EqXXVIII}) one sees that in the limit $e^\frac{\nu_2-\nu_1} {T} \rightarrow 0$ the distribution function $ \phi_{id} (k, T, \nu_2)$ with $\zeta \neq 0$ acquires  the Boltzmann  form.  In this limit  we find $p_{id}(T, \nu_2) \simeq p_{id}(T, \nu_1) \, e^\frac{\nu_2-\nu_1} {T}$ and $n^{(0)}_{id\,2} \simeq  \, n^{(0)}_{id\,1} \, e^\frac{\nu_2-\nu_1} {T}$. Using  these results and  Eq.  (\ref{EqXXIX}) we can rewrite  (\ref{EqXXIII}) as
\begin{eqnarray}
\label{EqXXXI}
&&\Sigma \simeq  R_p \left[  p_{id}(T, \nu_1)  \, e^\frac{ (1-\alpha)S_0\,  \Sigma} {T}   - P_{int\,2} (T,  n^{(0)}_{id\,2})  \right]  \, .\quad
\end{eqnarray}
Here we use the same notation as in previous section (see a paragraph before Eq.  (\ref{EqXIV})). From Eq. (\ref{EqXXXI})
one can see that  for    $\frac{V_0 \,p_{id}(T, \nu_1)} {T} \gg 1$  the surface tension coefficient $\Sigma$ is strongly suppressed compared to
$R_p\, p_{id}(T, \nu_1) $, i.e.  one finds
\begin{eqnarray*}
\Sigma \simeq  \frac{T} { S_0\, (\alpha-1) } \,   \ln \left[ \frac{R_p \, p_{id}(T, \nu_1)} {\Sigma }  \right] \ll R_p \, p_{id}(T, \nu_1) \,. 
\end{eqnarray*}
Note that for $\alpha >1$ the condition $e^\frac{\nu_2-\nu_1} {T} \rightarrow 0$  can be provided by $S_0 \Sigma/T  \gg 1$ only. Thus, the second term on the right hand side of  Eq. (\ref{EqXXXI}) cannot dominate, since it is finite. It is evident that  the inequality  $\frac{V_0 \,p_{id}(T, \nu_1)} {T} \gg 1$ also means that $n^{(0)}_{id\,1} V_0 \gg 1$, therefore,  in this limit  the effective chemical potential  (\ref{EqXXIV}) can be  approximated as
\begin{eqnarray}
\label{EqXXXII}
&&\nu_1 \simeq  \mu -  V_0 \, p + U_1(T, n^{(0)}_{id\,1}) \, ,
\end{eqnarray}
i.e. the contribution of the induced  surface tension is negligible  compared to the  pressure. This result means that  for
$n^{(0)}_{id\,1} V_0 \gg 1$, i.e. at high particle densities or for  $e^\frac{\nu_2-\nu_1} {T} \rightarrow 0$,  the IST EoS corresponds to the proper volume approximation.

On the other hand, Eq. (\ref{EqXXX}) exhibits that   at  low densities, i.e. for $e^\frac{\nu_2-\nu_1} {T} \simeq 1$,   the  IST EoS recovers the virial expansion of the gas of hard-spheres  up to fourth  power of  particle density $n_{1}$. Therefore, it is natural to expect that for intermediate values of the parameter $e^\frac{\nu_2-\nu_1} {T} \in [0; 1]$  the IST EoS will gradually  evolve from the
low density approximation to the high density one, if the condition  (\ref{EqXXIX}) is obeyed. This is a generalization of the previously obtained result \cite{VetaNEW} onto the  quantum statistics case.

Already from the virial expansion   (\ref{EqXXX}) one can see that the case $\alpha =1$ recovers the VdW EoS with the hard-core repulsion.  If, in addition,
the mean-field potentials are the same, i.e. $U_2 = U_1$ and, consequently, $P_{int\, 2} = P_{int\, 1}$, then  one finds that
$\nu_2 = \nu_1$ and $\Sigma = R_p\, p (T, \nu_1)$. In this case  the term $V_0\, p + S_0\, \Sigma \equiv 4 V_0\, p$ exactly corresponds to the VdW hard-core repulsion. If, however,   $U_2 \neq U_1$, but both mean-field potentials  are restricted from above, then  the model can deviate from the VdW EoS at low temperatures only, while at high temperatures it again corresponds to the VdW EoS.  For the case $U_2 < U_1$  this can be easily seen from  Eqs. (\ref{EqXXVII}) and  (\ref{EqXXVIII}) for the case $e^\frac{\nu_2-\nu_1} {T} \simeq 0$, if one sets $\alpha=1$. Then using the same logic as  in deriving Eq. (\ref{EqXXXI}), one can  find  that $\Sigma \ll  R_p \, p_{id} (T,\nu_1)$ and, hence, the effective chemical potential $\nu_1$ acquires the form  (\ref{EqXXXII}).  In other words, at low $T$ the surface tension effect becomes negligible and the IST EoS
corresponds to the proper volume approximation, if $e^\frac{\nu_2-\nu_1} {T} \simeq 0$.

Finally, if  the inequality  $U_2 >  U_1$ is valid, then at low $T$  an expansion  (\ref{EqXXVII}) has to be applied
to the distribution function $\phi_{id} (k, T, \nu_1)$
 instead of  $\phi_{id} (k, T, \nu_2)$ and then one arrives at the  unrealistic case, since $\Sigma \gg  R_p \, p_{id} (T,\nu_1)$.
 In this case  the hard-core repulsion would be  completely dominated by the induced surface tension term and, hence,
 even the second virial coefficient would not correspond to  the excluded volume of particles.

\section{Going beyond VdW approximation}

Let us closely inspect the IST EoS and show explicitly  its  major differences from  the VdW one.
For such a purpose in this section we  analyze  its effective and true  virial expansions and discuss
somewhat unusual properties of the entropy density.
\vspace*{-2.2mm}
\subsection{Effective virial expansion}
\vspace*{-2.2mm}
First we analyze the particle densities
$n_1 (T, \nu_1) \equiv  \frac{\partial p (T, \nu_1)} {\partial ~\mu}$ and  $\tilde n_2 (T, \nu_2) \equiv  R_p^{-1}\,\frac{\partial \Sigma (T, \nu_2)} {\partial ~\mu}$. For this purpose we differentiate Eqs.   (\ref{EqXXII}) and  (\ref{EqXXIII}) with respect to $\mu$
and apply the self-consistency conditions (\ref{EqXXVI})
\begin{eqnarray}
\label{EqXXXIII}
&&\hspace*{-4mm}n_1 = n_{id\,1} \left[  1 - V_0  n_1 - S_0 \frac{\partial \Sigma} {\partial \mu} \right] \,, \\
\label{EqXXXIV}
&&\hspace*{-4mm}
\frac{\partial \Sigma} {\partial \mu} = R_p \,   n_{id\,2} \left[  1 - V_0  n_1 - \alpha S_0 \frac{\partial \Sigma} {\partial \mu} \right]\,.
\end{eqnarray}
Expressing $\frac{\partial \Sigma} {\partial \mu}$ from Eq.  (\ref{EqXXXIV}) and substituting it into  (\ref{EqXXXIII}),
one finds the densities  of particle  number   ($ \tilde n_2 (T, \nu_2) \equiv n_2 (1- V_0 n_1)$)
\begin{eqnarray}
\label{EqXXXV}
&&\hspace*{-4.4mm}
n_1 =
\frac{n_{id\,1} \left(1 - 3 \, V_0  \, n_2 \right)} {1 + V_0  \, n_{id\,1}\left(1 - 3 \, V_0  \, n_2 \right)}\,,\\
\label{EqXXXVa}
&&\hspace*{-4.4mm} n_2  =
\frac{n_{id\,2}}{1 \, +\,   \alpha \, 3\,  V_0 \, n_{id\,2}}  \,, ~
\end{eqnarray}
where we used the relation $R_p S_0 = 3 V_0$ for hard spheres.
From Eq.  (\ref{EqXXXVa}) for $n_2$ one finds that for $\alpha > 1$ the term $(1 - 3 \, V_0  \, n_2)$ staying  above is always positive,
{ since, taking the limit  $n_{id\,2} \rightarrow \infty$ in  Eq.  (\ref{EqXXXVa})
one finds the limiting density of  $\max \{n_{2}\} = \left[ 3 \alpha V_0\right]^{-1}$.
}
Therefore,  irrespective of the  value  of  $n_{id\,2} \ge 0$  in the limit  $n_{id\,1} V_0 \gg 1$ one finds that $\max \{n_{1}\} = V_0^{-1}$.  This is  another  way to prove that the limiting density of the IST EoS corresponds to the proper volume limit, since at high densities it is four times higher than the one of the VdW EoS.
{Writing   the particle number density $n_{id\,1}$ from Eq.  (\ref{EqXXXV}) as
\begin{eqnarray}
\label{EqXXXVb}
&&
%%\hspace*{-8mm}
n_{id\,1} = \frac{n_1 }{\left(1 -  V_0  \, n_1 \right) \left(1 - 3 \, V_0  \, n_2 \right)}  \,, ~
\end{eqnarray}
one can get the formal virial-like expansion for the IST pressure $p_{id}(T, \nu_1  ) $ (\ref{EqXXII})
\begin{eqnarray}\label{EqXXXVc}
%%% \hspace*{-5.5mm}
\frac{p_{id} (T, \nu_1 )} {T } &=&
\sum\limits_{k=1}^\infty  \frac{a_k^{(0)} }{[1 - 3 \, V_0  \, n_2 ]^k} \frac{ \left[ n_1  \right]^{k} } {\left[1- V_0 \, n_1  \right]^k } \,,
\end{eqnarray}
where the expressions for the coefficients $a_k^{(0)}$ are given by Eqs. (\ref{EqIX})-(\ref{EqXVI}). This result allows us
to formally write an  expansion
\begin{eqnarray}\label{EqXXXVd}
%% \hspace*{-5.5mm}
\frac{p_{id} (T, \nu_1 )} {T } &\equiv&  \sum\limits_{k=1}^\infty  a_k^{(0), IST}  \frac{ \left[ n_1  \right]^{k} } {\left[1- V_0 \, n_1  \right]^k }
\end{eqnarray}
with the coefficients  $a_k^{(0), IST} = \frac{a_k^{(0)} }{[1 - 3 \, V_0  \, n_2 ]^k}$ which depend not only on $T$, but also on $n_2$.  
{The expansions  (\ref{EqXXXVc}) and (\ref{EqXXXVd})  are the generalizations of  the ones used for EoS of hard discs \cite{HardD}
and hard spheres \cite{HardS}.}

Similarly to deriving Eq. (\ref{EqXXIIe}), from (\ref{EqXXXVd})  one can get the quantum virial expansion  for IST  pressure  $p_{id} (T, \nu_1 ) $
\begin{eqnarray}\label{EqXXXVe}
%\hspace*{-5.5mm}
p_{id} (T, \nu_1 ) & =& T \sum\limits_{k=1}^\infty  a_k^{Q, IST}  n_1^k\,,\\
\label{EqXXXVf}
% \hspace*{-5.5mm}
a_k^{Q, IST} & =&   \sum\limits_{l=1}^{k} \frac{C_l^{(k)}}{[1 - 3 \, V_0  \, n_2 ]^l} \,,\\
\label{EqXXXVf2}
C_l^{(k)} & =&  \frac{(k-1)!}{(l-1)! (k-l)!}  V_0^{k-l} a_l^{(0)}  \,,
\end{eqnarray}
 with the coefficients $a_k^{Q, IST} $
which are $T$ and $n_2$ dependent.  For the interaction  pressure $P_{int\,1} (T,  n_{id\,1})$ which is expandable in terms of
the density $n_1$, Eq. (\ref{EqXXXVf}) can be used to estimate the full quantum virial coefficients of higher orders.
Of course, Eq. (\ref{EqXXXVe})   is not the traditional virial expansion, but the fact that it can be exactly obtained  from the grand canonical ensemble formulation of  the quantum version of the IST EoS
for the third, the fourth and higher order virial coefficients is still remarkable.
%%%%%
\vspace*{-2.2mm}
\subsection{True quantum virial coefficients}
\vspace*{-2.2mm}
{Now we consider an example on how to employ the results  (\ref{EqXXXVe})-(\ref{EqXXXVf2}) to estimate the
true virial coefficients at low densities and at sufficiently high temperature which provide the convergence of virial expansion (\ref{EqXXXVe}). Apparently, in this case one can expand the density $n_2 \simeq B_1 n_1  (1 + B_2 n_1)$ in powers of the density $n_1$.  From our above  treatment of the low density limit $e^\frac{\nu_2-\nu_1} {T} \simeq 1$ it is clear that $B_1 =1$.
Substituting  this  expansion for $n_2$ into Eqs.  (\ref{EqXXXVe}) and (\ref{EqXXXVf}) and keeping only the terms up to $n_1^2$ one can get  the true quantum virial coefficients  $a_k^{Q, tot}$ as
\begin{eqnarray}
\label{EqXXXVf3}
%%% \hspace*{-4.4mm}
&&a_{2}^{Q, tot} =   V_0 + a_2^{(0)} + 3V_0 B_1 = 4V_0 + a_2^{(0)} \,,\\
 \label{EqXXXVf4}
 %
 %% \hspace*{-4.4mm}
&&a_{3}^{Q, tot}  \simeq  13 V_0^2 + 3 V_0 B_2 +  5 V_0 a_2^{(0)} + a_3^{(0)} \,,\\
  %
%%   \hspace*{-4.4mm}
 \label{EqXXXVf5}
&&a_{k\ge3}^{Q, tot}  \simeq    \sum\limits_{l=1}^{k} C_l^{(k)} + 3V_0 B_1 \sum\limits_{l=1}^{k-1} C_l^{(k-1)} l \nonumber  \\
&&+ 3V_0 B_1 \sum\limits_{l=1}^{k-2} C_l^{(k-2)} \left[ \frac{3}{2}l(l+1) V_0 B_1 +B_2 \right] \,,~
\end{eqnarray}
and replace the coefficients $a_k^{Q, IST}$ in Eq.  (\ref{EqXXXVe})  with the true quantum virial coefficients $a_{k}^{Q, tot}$ which depend on $T$ only.
Note that  an expression  for the second virial coefficient  $a_{2}^{Q, tot}$  is exact, while the  expressions for the higher order virial coefficients are the approximated ones, which, nevertheless,  at high values of temperature are rather accurate.
Considering the limit of high temperatures which allows one to  ignore  the  quantum corrections in Eqs. (\ref{EqXXXVf3})
and  (\ref{EqXXXVf4}), one can find  the coefficients $B_1 =1$ exactly and $B_2 \simeq [7 -6\alpha]V_0$ approximately by comparing the expressions
 (\ref{EqXXXVf3}) and  (\ref{EqXXXVf4})  with the corresponding virial coefficients of Boltzmann gas in Eq.  (\ref{EqXXX}).
 Substituting the obtained expressions  for  $B_1$ and $B_2$ coefficients  into Eq.  (\ref{EqXXXVf5}) one gets the approximate  formula for higher order virial coefficients $a_{k\ge3}^{Q, tot}$:
\begin{eqnarray}
%
%%%   \hspace*{-4.4mm}
 \label{EqXXXVf6}
&&\hspace*{-3.3mm}a_{k\ge3}^{Q, tot}  \simeq    \sum\limits_{l=1}^{k} C_l^{(k)} + 3V_0  \sum\limits_{l=1}^{k-1} C_l^{(k-1)} l \nonumber  \\
&&\hspace*{-3.3mm}+ 3V_0^2  \sum\limits_{l=1}^{k-2} C_l^{(k-2)} \left[ \frac{3}{2}l(l+1)   + (7-6\alpha) \right] = \nonumber  \\
&&\hspace*{-3.3mm}=\sum\limits_{l=1}^{k}  \frac{(k-1)!  V_0^{k-l} a_l^{(0)}}{(l-1)! (k-l)!} + 3  \sum\limits_{l=1}^{k-1} \frac{(k-2)!  V_0^{k-l} a_l^{(0)}}{(l-1)! (k-1-l)!}  l  ~ \nonumber  \\
&&\hspace*{-3.3mm}+3 \sum\limits_{l=1}^{k-2}  \frac{(k-3)!  V_0^{k-l} a_l^{(0)}}{(l-1)! (k-2-l)!}  \left[ \frac{3}{2}l(l+1)   + (7-6\alpha) \right] 
\,,~
\end{eqnarray}
where the  second equality above is obtained by substituting  Eq. (49) for the coefficients $C_l^{(k)}$ into  the first one.

Comparing  now Eq. (\ref{EqXXXVf6}) for the IST EoS   and  Eq. (\ref{EqXXIIe}) for  the VdW EoS one can see that the first sum on the
right hand side of (\ref{EqXXXVf6})  is identical to the expression for the VdW quantum virial coefficients with the excluded volume $b = 4 V_0$
 replaced by the proper  volume $V_0$.  Apparently, the other two sums on the  right hand side of (\ref{EqXXXVf6}) are the corrections
 due to induced surface tension coefficient.

 Note  that
it is not difficult to  get the exact expressions for the third or the fourth virial coefficients  $a_{k}^{Q, tot}$
by inserting the  higher order terms of  the expansion  $n_2 (n_1)$  in power of  density $n_1$ into Eqs. (\ref{EqXXXVe}) and (\ref{EqXXXVf}), although comparing the coefficients in front of  $B_1$ and $B_2$ in the last sum of  Eq.  (\ref{EqXXXVf5}),
one can see that even for $l=1$ the coefficient staying before $B_1$ is essentially larger than the one staying  before $B_2$.
This means that  at low densities  the role of  $B_2$ is an auxiliary one, if $\alpha$ is between 1 and 1.5.
}
%%%%%
\vspace*{-2.2mm}
\subsection{Virial expansion for compressible spheres}
\vspace*{-2.2mm}
It is interesting that the $k$-th term 
\begin{eqnarray*}
\frac{1 }{[1 - 3 \, V_0  \, n_2 ]^k} \frac{ \left[ n_1  \right]^{k} } {\left[1- V_0 \, n_1  \right]^k } \,,
\end{eqnarray*}
staying 
in the  sum  (\ref{EqXXXVc}) allows for a non-trivial interpretation.  Comparing Eq. (\ref{EqXVII}) and  Eq. (\ref{EqXXXVc}) and recalling the fact that the particle number density $n_1$ is proportional to the number of spin-isospin configurations $ d_p$, one can introduce an effective number of
such configurations as $d_p^{eff} = \frac{d_p}{1- 3 V_0 n_2}$ with simultaneous replacement of $V_0$ by  the effective proper volume
 \begin{eqnarray*}
 V_0^{eff} = V_0\, (1- 3 V_0 n_2) 
\end{eqnarray*}
 in all terms which contain the powers of   $[1 -V_0 n_1]$   on the right hand side of   (\ref{EqXXXVc}).  Then at high densities  the   effective number   of spin-isospin configurations $d_p^{eff} \le  \frac{\alpha \, d_p}{\alpha -1}$ can be sizably larger than $d_p$, while
the effective proper volume $V_0^{eff}$ can be essentially smaller than  $V_0$ (i.e. such effective particles are compressible), if the coefficient $\alpha > 1$  is close to 1.
Moreover, one can also establish an equivalent virial expansion of  pressure (\ref{EqXXXVc}) in terms of $\frac{n_1}{(1- 3 V_0 n_2)}$ powers. Then instead of  the coefficients $a_k^{Q, IST} $ (\ref{EqXXXVf}) one would get
\begin{eqnarray}
\label{EqXXXVg}
% \hspace*{-5.5mm}
\tilde a_k^{Q, IST} & =&   \sum\limits_{l=1}^{k} \frac{(k-1)!}{(l-1)! (k-l)!}  \left[V_0^{eff} \right]^{k-l}a_l^{(0)}  \,,
\end{eqnarray}
which shows that at high densities  the contributions  of low order virial coefficients $a_l^{(0)}$ into the coefficient
$\tilde a_{k>1}^{Q, IST}$ are  suppressed  due to decrease of $V_0^{eff} $.   Eq.  (\ref{EqXXXVg}) quantifies the
source of softness of the IST EoS compared to VdW one at high densities.
 It is also interesting
that the monotonic  decrease of  $V_0^{eff} $ at high densities is qualitatively similar to the effect of Lorentz contraction of proper volume for  relativistic  particles \cite{RelVDW}.

Although  the present model
does not   know anything about the internal structure of considered  particles, but the fact that $d_p^{eff}$ increases with the particle number density $n_2$ can be an illustration of the in-medium effect   that the IST  hard-core interaction `produces'
  the additional (or `enhances' the number of existing) spin-isospin states which  are well known  in quantum physics  as excited states, but with an  excitation energy
being essentially smaller than  the mean value of  particle free energy.  In this way one can see  that at high densities
 the IST  effectively increases  the degeneracy factor  of particles.
This finding is  a good illustration that   the claim of  Ref.  \cite{Typel16} that
 accounting for the excluded volume correction in quantum case via  the effective degeneracy  leads to the reduction of latter
 (see a discussion of  Eqs.(18) and (19) in \cite{Typel16}) is not a general one.
{On contrary,  a more advanced EoS developed above requires not a reduction of the effective number of degrees
of freedom as it is suggested in \cite{Typel16}, but to their enhancement. 
}

It is apparent that for $\alpha \gg 1$ the quantities   $V_0^{eff} $ and  $d_p^{eff}$ are practically independent of $n_2$, i.e.
in this case the coefficients $a_k^{Q, IST}$ and  $\tilde a_k^{Q, IST}$ are the true quantum virial coefficients of the VdW EoS, but with the excluded volume $b = 4 V_0$ replaced by  $V_0$.
}
%%%%%
\vspace*{-2.2mm}
\subsection{Properties of entropy density}
\vspace*{-2.2mm}
Next we study the entropy density of the IST EoS. Similarly to finding the derivatives of  Eqs.   (\ref{EqXXII}) and  (\ref{EqXXIII}) with respect to $\mu$, one has to find their derivatives with respect to $T$ in order to get
the entropy per particle
\begin{eqnarray}
\hspace*{-8mm} \label{EqXXXVI}
&&\frac{s_1} {n_1} = \frac{
\left[ \frac{\tilde s_{id\,1} } {n_{id\,1}} - 3 \, V_0  \, n_2 \cdot \frac{\tilde s_{id\,2}} {n_{id\,2}} \right]
}{ \left[1 -  3 \, V_0  \, n_2 \right]  } \,,  \\
\label{EqXXXVII}
\hspace*{-4mm}
&&\tilde s_{id\,A} \equiv s_{id\,A} + n_{id\,A} {\frac{\partial U_A } {\partial ~ T} -  \frac{\partial P_{int\,A} } {\partial ~ T} } \,,
\end{eqnarray}
where the entropy density of point-like particles is defined as
$s_{id\,A} \equiv \frac{\partial p_{id} (T, \nu_A)} {\partial ~T}$ and $A \in \{1; 2\}$. If the mean-field potentials of the model have the form
\begin{eqnarray}\label{Eq50VII}
U_A = \sum_\lambda g^\lambda_A(T) f^\lambda_A(n_{id\,A})
\end{eqnarray}
and for $T = 0$ their derivatives obey the set of  conditions  $\frac{d g^\lambda_A(T)} {d ~T} = 0$, then it is easy to see that the entropy per particle $\frac{s_1} {n_1}$ also vanishes at $T=0$, i.e. the Third Law of thermodynamics is obeyed under these conditions.
 In a special case, when interaction mean-field potentials do not explicitly  depend on
the temperature $T$ an expression for the entropy densities (\ref{EqXXXVII}) gets simpler, i.e. $\tilde s_{id\,A} = s_{id\,A}$. This
case is  important for the hadron resonance model and, hence,  it is discussed in the Appendix in some details.

Apparently, to provide a positive value of entropy per particle $\frac{s_1} {n_1}$ one has to properly choose the interaction terms in Eqs.  (\ref{EqXXII}) and  (\ref{EqXXIII}). 
In other words, the  Third Law of thermodynamics provides one of the basic constraints to the considered EoS.
 It is clear that the  corresponding necessary conditions
should not be very restrictive because at low densities, i.e.  for $3 \, V_0  \, n_{2} \ll 1$, the coefficient staying in front  of the term  $\frac{\tilde s_{id\,2}} {n_{id\,2}}$ is very small, while at high densities it is $\alpha^{-1}< 1$ for  $\alpha > 1$.
Although,  a discussion of  such conditions is far beyond the scope  of this work,  below we consider two important cases.

For the case $U_2(T, \rho) \equiv U_1(T, \rho)$  the condition (\ref{EqXXIX}) is valid for any choice
of parameters. Then
one can  show  a validity of the inequality $\frac{s_{id\,1}} {n_{id\,1}} \ge \frac{s_{id\,2}} {n_{id\,2}}$, since for $\alpha >1$ one finds $\nu_1 > \nu_2$.  To prove this inequality one has to account that  $p_{id} (T, \nu_A)$ and all its derivatives are monotonously increasing functions of $T$ and $\nu_A$.  Then, using the relations  (\ref{EqXXVII}) and  (\ref{EqXXVIII})  between  the quantum distribution functions,  one can show the validity of the inequality  $\frac{s_{id\,1}} {n_{id\,1}} \ge \frac{s_{id\,2}} {n_{id\,2}}$ for two limits $e^\frac{\nu_2-\nu_1} {T} \simeq 1$
and $e^\frac{\nu_2-\nu_1} {T} \rightarrow 0 $. Similarly, one can introduce an
effective parameter
of statistics 
\begin{eqnarray*}
\zeta_{eff} \equiv \zeta - \zeta \left[1 - e^\frac{\nu_2-\nu_1} {T}  \right]  
\end{eqnarray*}
and study the quantities for the distribution function $\phi_{id} (k, T, \nu_2) $ with an effective parameter of statistics $\zeta_{eff}$. However, one can easily understand that the inequality
$\frac{s_{id\,1}} {n_{id\,1}} \ge \frac{s_{id\,2}} {n_{id\,2}}$  cannot be broken down 
for any value of  the exponential   $e^\frac{\nu_2-\nu_1} {T}$ obeying the inequalities
  $0< e^\frac{\nu_2-\nu_1} {T} < 1$. This is so,  since  the pressure  of point-like particles and its
partial  derivatives are monotonic functions of the parameters $T$ and $\nu_1$ (or $\nu_2$)  and that a non-monotonic behavior of the entropy per particle can be caused by the phase transition, which does not exists for an ideal gas.  Note that  here we do not consider a possible effect
of the Bose-Einstein condensation. Using the above inequality  between the entropies per particle and requiring that  $U_1 \ge 0$ and the inequalities  $\frac{d g^\lambda_A(T)} {d ~T} > 0$ for $T>0$ and $\frac{d g^\lambda_A(T=0)} {d ~T} = 0$ one can show that  $\frac{s_{1}} {n_{1}} \ge \frac{\tilde s_{id\,2}} {n_{id\,2}} \ge 0$ using an  identity (\ref{EqV}).

Another important case corresponds to the  choice $U_1 > 0$ and $U_2 < 0$ in Eq. (\ref{Eq50VII}), i.e. the mean-field $U_1$ describes an attraction, while  $U_2$ represents a repulsion. Clearly, the condition  (\ref{EqXXIX}) in this case is also fulfilled for  any choice of parameters.
 Using the self-consistency relations (\ref{EqXXVI}), or its more convenient  form  (\ref{EqV}), one can find that the  term describing the entropy of mean-field  in $\tilde s_{id\,2}$  can be negative, i.e.  
 \begin{eqnarray}
n_{id\,2} \frac{\partial U_2 } {\partial ~ T} -  \frac{\partial P_{int\,2} } {\partial ~ T} =
\sum_\lambda \frac{d   g^\lambda_2 (T)} {d ~ T}  \int\limits_0^{n_{id\,2}} \hspace*{-1.1mm} d n \,  f ^\lambda_2 (n) < 0 \,, ~
\end{eqnarray}
if $g^\lambda_2 (T) > 0$,   $\frac{d   g^\lambda_2 (T)} {d ~ T}  > 0$, but  $U_2 < 0$  for $T \ge  0$
due to the inequalities $ f ^\lambda_2 (n) < 0$. 
Such  a choice of interaction allows one to decrease the effective entropy density  $\tilde s_{id\,2}$ or even to  make it negative by tuning the mean-field  $U_2$ related to the IST coefficient.  As a result this would increase the physical entropy density $s_{1}$.  Note that for the VdW EoS this is impossible.

 \section{Application to nuclear and hadronic matter} 
 \subsection{Some important  examples}
 As a pedagogical example to our discussion  we consider the IST EoS for the nuclear matter and compare it with the VdW EoS  (\ref{EqI}) having the following interaction
\vspace*{-2.2mm}
\begin{eqnarray}\label{EqXXXVIII}
%\hspace*{-7mm}
&& P_{int}^{VdW} (T,  n_{id})  = a \left[ \frac{n_{id}} {1 + b\, n_{id}} \right]^2 + T n_{id} -
\frac{g(T) \,n_{id}} {1 + b\, n_{id}}
\nonumber   \\
\hspace*{-7mm}&&  - \frac{g(T)  b\,n^2_{id}} {\left[1 + b\, n_{id} \right]^2} - \frac{g(T) \, B_3\, n_{id}^3} {\left[1 + b\, n_{id} \right]^3} - \frac{g(T) \, B_4\, n_{id}^4} {\left[1 + b\, n_{id} \right]^4}  \,, \quad \quad
\end{eqnarray}
where the virial coefficients $b$,  $B_3$ and  $B_4$ are introduced above and the function $g(T) \equiv \frac{T^2} {T + T_{SW}}$ with $T_{SW}=1$ MeV provides the fulfillment of the Third Law of thermodynamics.  Note that the term $T n_{id}$ cancels exactly the first term of the quantum virial expansion for $p_{id} (T,\nu)$ (see Eq. (\ref{EqXVII})),
while the  term $a \left[ \frac{n_{id}} {1 + b\, n_{id}} \right]^2 $ in Eq.  (\ref{EqXXXVIII}) accounts  for an attraction and the other terms proportional to $g(T)$  are the lowest  four  powers of the virial expansion for the gas of classical hard spheres for $T \gg  T_{SW}$.  By construction, such an EoS, apparently, reproduces the four first virial coefficients of the gas of hard spheres
at $T \gg  T_{SW}$ and, simultaneously, it obeys the Third Law of thermodynamics at $T=0$.
\begin{figure}
[ht]
\vspace*{-0.0mm}
\centerline{\includegraphics[width=86mm]{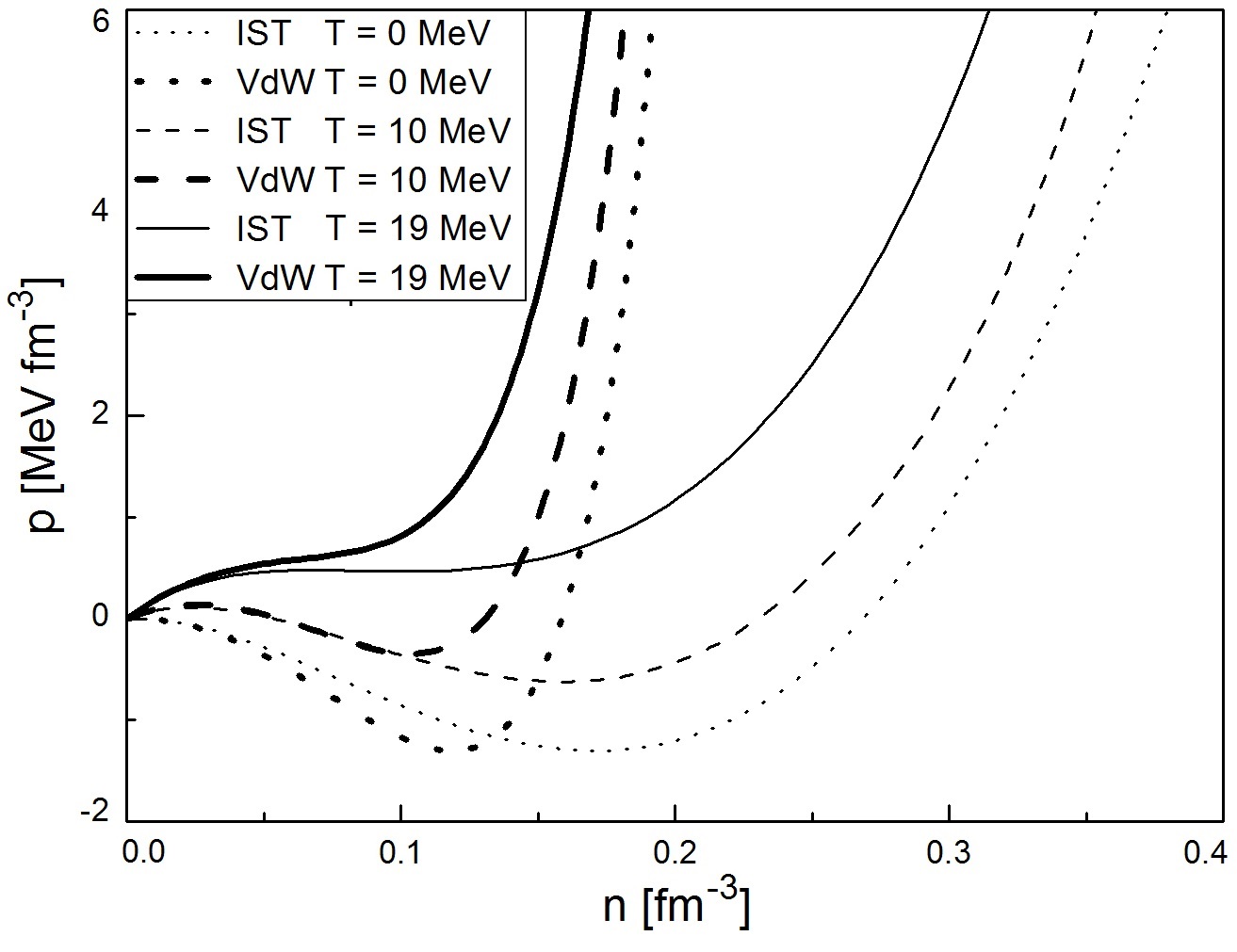}}
\vspace*{-5.5mm}
 \caption{Behavior of  pressure as a function of particle number density for isotherms of nuclear matter (see text for details).
 }
  \label{Fig1}
\end{figure}

For the IST EoS we choose $\alpha =1.245$ \cite{VetaNEW},  $P_{int\,1}^{IST}=a \left[ \frac{n_{id\,1}} {1 + b\, n_{id\,1}} \right]^2$ and $P_{int\,2}^{IST}=0$ with the same  constants $a \simeq 329$ MeV fm$^3$ and $b = 4 V_0 \simeq 3.42$ fm$^3$ which  were found in   \cite{Vovch15} for VdW EoS of  nuclear matter $(d_p=4, m_p= 939 {\rm \, MeV})$, i.e. we took just the parameters of Ref.  \cite{Vovch15} for a proper comparison.
By construction the IST EoS  and  EoS  (\ref{EqXXXVIII}) agree very well (within one  percent)  for $T> 120$ MeV and particle number densities $n\le 0.25$ fm$^{-3}$.
{In Fig. 1 we compare  three isotherms $T = 19, 10$ and 0 MeV of these two EoS.
For $T=10$ MeV their  isotherms  agree up to the  packing fraction $\eta = V_0 n \simeq 0.09$ (for the nuclear density $n \le 0.11$ fm$^{-3}$), i.e. within the usual range of the VdW EoS applicability \cite{Bugaev:ALICE,VetaNEW}.
However,
for $T=0$ and $T=19$ MeV  isotherms the both models agree up to  the packing fraction $\eta = V_0 n \simeq 0.03$ only (for  $n \le 0.035$ fm$^{-3}$), i.e. far below  the usual range of the VdW EoS applicability   due to  important  role of the second and higher  order quantum virial coefficients $a_{k\ge2}^{(0)}$ defined by Eqs. (\ref{EqX})-(\ref{EqXV}).
The present example clearly shows that providing the four virial coefficients of the gas of hard spheres  for the  quantum VdW EoS of Ref. \cite{Vovch15}  at high temperatures, one can, at most, get a good agreements with the IST EoS for a single value of temperature, namely for  $T=10$ MeV.   Fig. 1 also shows that
for the same parameters the IST EoS is essentially softer that the improved VdW one, hence, it does not require so strong attraction
and so strong repulsion to reproduce the properties of normal nuclear matter. This conclusion is supported by the results obtained recently  for
nuclear matter EoS  within the  IST  concept  \cite{Aleksei17}.}

Recently an interesting
generalization of the quantum VdW EoS (GVdW hereafter)    was suggested in \cite{Vovch17}. This  EoS  allows one to go beyond the VdW approximation, but  formally it is similar to the VdW models discussed above.  In terms of  the ideal gas pressure (\ref{EqII}) the GVdW pressure can be written  as  \cite{Vovch17} ($\eta = V_0 n$ is the packing fraction):
\begin{eqnarray}\label{EqXXXIX}
%
%%\hspace*{-7mm}
&&p_G ( T, \mu)  = w (\eta) \, p_{id} (T, \nu_G) - P_{int\,G} ( n ) \, , \quad\\
\label{EqXXXX}
%
%%\hspace*{-7mm}
&& \nu_G (\mu, n) = \mu + V_0\, f^\prime(\eta) \, p_{id}(T, \nu_G)  + U_G (n) \, ,
\end{eqnarray}
where
%%the packing fraction $\eta = V_0 n$ is given in terms of particle density $n$,
$n$ is the particle density, and
the multiplier  $w (\eta) \equiv (f(\eta) -\eta f^\prime(\eta))$  is given in terms of the function $f(\eta)$ which 
is defined as
\begin{eqnarray}\label{60II}
{f(\eta)} =
 \left\{
\begin{array} {lll}
f^{VdW} (\eta)= {1- 4 \eta}  \, ,  &  {\rm for ~ VdW ~EoS}\,,   \\
& \\ 
f^{CS} (\eta) =\exp \left[ - \frac{(4-3\eta)\eta} {(1-\eta)^2} \right]  \, ,  &  {\rm for ~ CS ~EoS}  \,,
\end{array}
\right.
\end{eqnarray}
where the function $f^{VdW} (\eta)$ corresponds to the VdW case,
whereas the function $f^{CS} (\eta)$ is given for the famous Carnahan-Starling (CS) EoS \cite{CSEoS}.
 The interaction terms  of the GVdW EoS are given  in terms of a function $u(n)$: $U_G = u(n) + n  u^\prime (n)  $ and  $P_{int\,G} = -  n^2 u^\prime (n)$.  This  choice automatically provides  the  self-consistency condition fulfillment.
Since the potentials $U_G$ and $P_{int\,G}$
are temperature independent, the Third Law of thermodynamics is obeyed.
\begin{figure}
[th]
\vspace*{-0.0mm}
\centerline{\includegraphics[width=86mm]{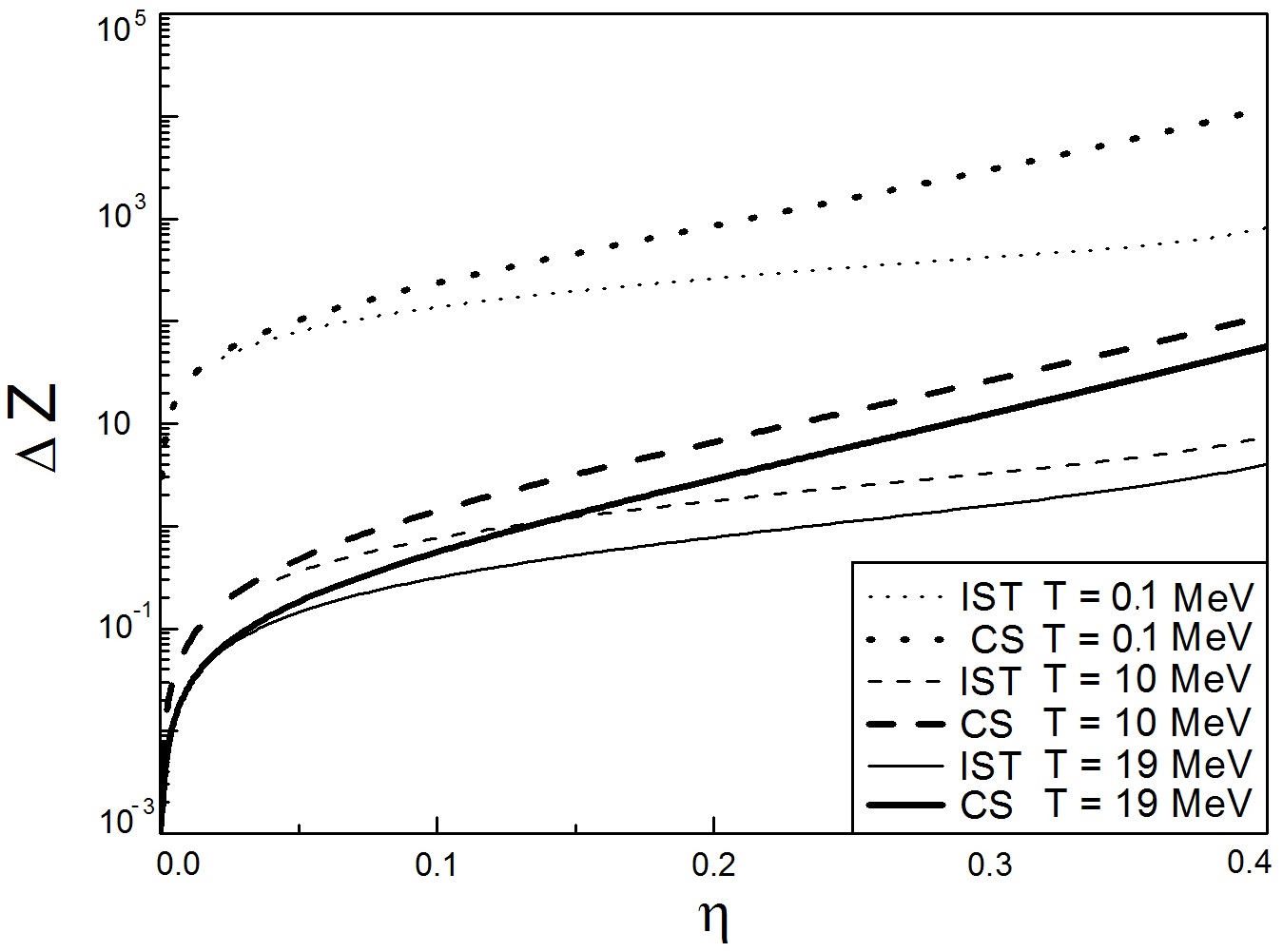}}
\vspace*{-4.4mm}
 \caption{Packing fraction dependence of   the quantum compressibility factors $\Delta Z_{Q}$ of the
 GVdW EoS and  IST EoS (see  text).}
\label{Fig2}
\end{figure}

The presence  of  the  function $w(\eta)$ in front of the ideal gas pressure in  (\ref{EqXXXIX})  allows one to reproduce the famous  CS EoS \cite{CSEoS} at high temperatures, while it creates the problems with formulating the  GVdW model for several hard-core radii, since the pressures of point-like particles of kinds 1 and 2 cannot be added to each other, if  their functions $w(\eta_1)$ and $w(\eta_2)$ are not the same.

Using the quantum virial expansion  (\ref{EqVIII}) and the particle number  density expression $n = f (\eta) \, n_{id}(T, \nu_G) $ \cite{Vovch17}, for $P_{IG} \equiv  w(\eta)  \,p_{id} (T, \nu_G )$  one obtains
%
%\vspace*{-2.2mm}
\begin{eqnarray}\label{EqXXXXI}
\hspace*{-8mm}&&P_{IG}  = w (\eta)  \, T \biggl[  \frac{n} {f(\eta)}   +  \sum\limits_{l=2}^\infty  a_l^{(0)} \left[  \frac{n} {f(\eta)}  \right]^l \biggr] \,, ~\quad\\
\label{EqXXXXII}
%
%\hspace*{-8mm}
&& \frac{w (\eta)} {f(\eta)} =
 \left\{
\begin{array} {lll}
\frac{1} {1- 4 \eta} \equiv \frac{1} {f^{VdW} (\eta)} \, ,  &  {\rm for ~ VdW ~EoS}\,,   \\
 %&   \\
%
\frac{1+ \eta + \eta^2 -  \eta^3} {(1- \eta)^3}  \, ,  &  {\rm for ~ CS ~EoS}  \,.
\end{array}
\right.
\end{eqnarray}
Although this effective expansion can be used to derive the true virial expansion for the CS parameterization of the GVdW  EoS (for the VdW one it is given above),
the result is  cumbersome.  Nevertheless,
these equations show that due to the multiplier $w (\eta)$ the first term of the quantum virial expansion  in Eqs. (\ref{EqXXXXI}),  (\ref{EqVIII}), (\ref{EqXVII}) and (\ref{EqXXXVe}), i.e. the classical term,   exactly reproduces the pressure of  corresponding classical EoS.
Hence, all   other  terms in  Eqs. (\ref{EqVIII}), (\ref{EqXVII}),  (\ref{EqXXXVe}) and (\ref{EqXXXXI})
are   the quantum ones.
A direct comparison of the IST with $\alpha=1.245$ and CS EoS for classical gases shows that for  packing fractions $\eta > 0.22$ the IST EoS is softer than the CS one \cite{Bugaev:ALICE,VetaNEW}.
 Fig.\ \ref{Fig2} depicts  the quantum compressibility factors 
\begin{eqnarray*}
\Delta Z^{CS}_{Q} (\eta) = \frac{P_{IG} - w(\eta) T n_{id} (T, \nu_G) } {T \, n}
\end{eqnarray*}
for the
CS EoS  of the GVdW model  and the one  for the IST EoS  defined similarly 
\begin{eqnarray*}
\Delta Z^{IST}_{Q} (\eta) = \frac{p_{id\, 1} -  T n_{id\, 1} (T, \nu_1) } {T \, n_1}
\end{eqnarray*}
taken both  for  the same  parameters $b=3.42$ fm$^3$, $P_{int\, G} (T,  n ) = a_{attr} n^2$ with $a_{attr} =329 $ MeV$\cdot$fm$^3$  (see \cite{Vovch17} for more details). 
As one can see from  Fig.\ \ref{Fig2} the quantum compressibility factors  of these EoS   differ essentially  for $\eta \ge 0.05$. Therefore, for $\eta \ge 0.1$ both the classical and the quantum parts of the IST  pressure   with $\alpha =1.245$ \cite{VetaNEW} are essentially softer than the corresponding terms of the CS version of  GVdW model  of  Ref. \cite{Vovch17}.
One can easily understand such a conclusion comparing the expansions (\ref{EqXXXXI}) and  (\ref{EqXXXVc}).
Since for the same packing fraction $\eta \ge 0.1$  the function $f^{CS}(\eta)$ of the CS version of  GVdW EoS  vanishes essentially faster than the term $[1-3 \,V_0 n_2][1-V_0 n_1]$ of the IST EoS, then each term proportional to $n^k$  in (\ref{EqXXXXI}) with $k> 1$  is larger than
the corresponding  term proportional to $n_1^k=n^k$ in (\ref{EqXXXVc}).  It is necessary to note  that such a property is very important  because
 the softer EoS provides a wider range of thermodynamic parameters for which the EoS is causal, i.e. its speed of sound is smaller than
the speed of light.

 \subsection{Constraints on nuclear matter properties}
It is appropriate to  discuss here  the most important   constraints on the considered  mean-field models which are necessary to describe the strongly interacting matter properties.  According to Eqs. (\ref{EqXVII}),  (\ref{EqXXXVe}) and  (\ref{EqXXXXI}) the
%%%repulsive part of
 fermionic  pressure of considered EoS consists of three contributions: the classical pressure  (the first term on the right hand side of    (\ref{EqXVII}),  (\ref{EqXXXVe}) and  (\ref{EqXXXXI})),  the quantum part of pressure
and
%%%the repulsive part of
 the mean-field $P_{int}$. At  temperatures below 1 MeV the classical part is negligible,
but
the usage of virial expansions discussed above  is   troublesome   due to convergency problem.

 Since
the exact  parameterization  of  the function  $P_{int}$ on the particle number density of nucleons is not known, it is evident that   all considered models are effective by construction. To fix their  parameters  one has  to reproduce the usual properties of normal nuclear matter, i.e. to get  a zero value for the  total pressure at normal nuclear density $n_0 \simeq 0.16$ fm$^{-3}$
and the binding energy $W= -16$ MeV at this density \cite{GenRevEoS}.
Similarly to high temperature case discussed at the end of Section 2 there is exist a freedom of parametrizing the
hard-core radius of nucleons, since the attraction pressure can be always adjusted to reproduce  the properties of normal nuclear matter and, therefore,  all the model parameters are also  effective by construction.

However, in addition to the properties of normal nuclear matter   there is the so called  flow   constraint at nuclear densities $n = (2-5) n_0$ \cite{Pawel}, which  sets strong restrictions
on  the model pressure  dependence on nuclear particle density and requires  rather soft  EoS  at  these densities. Hence, it can be used to determine the parameters of realistic EoS at high  nuclear densities and $T=0$.
Traditionally, such a constraint  creates  troubles for the relativistic mean-field EoS based on Walecka model \cite{Walecka74, Dutra14,Menezes17}. 

The validity  of this statement  can be seen   from Ref.   \cite{Dutra14} in which it is shown that 
  only 104  of such {EoSs} out of 263 analyzed  in   \cite{Dutra14}  are able to obey the flow constraint despite the fact that they have
10 or even more adjustable parameters.
At the same time as one can see from the simplest realization of the IST EoS suggested in   Ref.  \cite{Aleksei17},
the 4-parametric EoS is able to simultaneously  reproduce all properties of normal nuclear  matter and the flow constraint.
Furthermore, the  IST EoS  is able not only to reproduce the  flow constraint, but simultaneously  it is  able to
 successfully  describe  the  neutron star properties  with the masses more  than two Solar ones \cite{NS17}, which set
 another  strong constraint on the stiffness of the realistic EoS at high particle densities and zero temperature.
On the other hand, Fig. 2 shows that the existing CS version of  GVdW  EoS  of Ref.  \cite{Vovch17} is very stiff and, hence, it will also  have  troubles to obey the flow constraint  \cite{Pawel}.

 \subsection{Constraints on hadronic matter properties}
From the virial expansions of all models discussed  here one sees that the EoS  calibration on the properties of nuclear matter  at low T and  at  high densities involves mainly the quantum and the mean-field pressures, but, unfortunately, it also fixes the parameters of the  classical  pressure at higher temperatures. It is, however,  clear that the one component  mean-field models of nuclear matter cannot be applied at temperatures above 50 MeV, since one has to include the mesons, other baryons  and their resonances  \cite{Raju,Kostyuk00}. 

Moreover, at high temperatures  the mean-fields and the parameters of interaction  should be re-calibrated  because the very fact of resonance existence already corresponds to a partial accounting  of  the  interaction \cite{Raju}.
For many years it is well known  that for temperatures below 170 MeV and densities below $n_0$ the  mixture of stable hadrons and their resonances whose interaction is taken into account by the quantum  second virial coefficients
 behaves as the mixture of nearly ideal gases of stable particles which, in this case, includes both the hadrons and the resonances,  but  taken with their averaged masses  \cite{Raju}.  The main reason for  such a behavior is rooted in a nearly complete cancellation between the attraction and repulsion contributions. The resulting deviation from the ideal gas (a weak repulsion) is usually described in  the hadron resonance gas model (HRGM) \cite{Andronic06,Horn,KABOliinychenko:12,Bugaev13,Stachel:2013zma,Veta14,Bugaev:ALICE,VetaNEW,Bugaev17b} by   the classical  second virial coefficients.
 
Nevertheless, such a repulsion is of principal importance for the HRGM, otherwise, if one considers the mixture of  ideal gases of  all known hadrons and their resonances, then at high temperatures  the pressure of such a system will exceed the one of the ideal gas of massless quarks and gluons \cite{Satarov10}. Since such a behavior contradicts to the lattice version of quantum chromodynamics, the (weak) hard-core repulsion in  the HRGM is absolutely  necessary.
Moreover, to our best knowledge there is no other approach which is able to include all known hadronic states  into consideration and to be  consistent with the  thermodynamics of lattice quantum chromodynamics at low energy densities and which, simultaneously,  would   not contradict  it at  the higher ones.

Therefore,  it seems that  the  necessity  of  weak repulsion  between the hadrons   is naturally encoded in the smaller values of their hard-core radii ($R_p < 0.4$ fm) obtained within the HRGM  compared to  the larger hard-core radius
of nucleons in nuclear matter $R_n \ge 0.52$ fm  found in  \cite{Vovch17}.
This conclusion is well supported by the  recent simulations of the  neutron star properties
with masses more than two Solar ones \cite{NS17}
which also favor the  nucleon hard-core radii below than $0.52$ fm.
Furthermore, the  small values of the hard-core  radii provide the fulfillment of the causality condition   in hadronic phase \cite{Bugaev:ALICE,VetaNEW,NS17,Satarov15}, while a possible break of causality occurs in the region where the hadronic degrees of freedom  are not relevant \cite{Satarov15}.
 Hence, in contrast to Ref. \cite{Vovch17}, we do not see any reason to believe that the  mean-field model describing the nuclear matter properties may set any strict conditions on the hadronic hard-core radii of  the HRGM.

Moreover, we would like to point out  that a great success achieved recently  by  the  HRGM  \cite{Andronic06,Horn,KABOliinychenko:12,Bugaev13,Stachel:2013zma,Veta14,Bugaev:ALICE,VetaNEW,Bugaev17b} sets a strong restriction  
on any model of hadronic phase which is claimed  to be  realistic. 
The point is that  at the chemical freeze-out curve $\mu=\mu_{CFO}(T)$ the mean-field interaction term of pressure (\ref{EqI}) or (\ref{EqXXII}) must vanish, otherwise one would need a special procedure to  transform the
mean-field potential energy  into the masses and  kinetic energy of  non-interacting hadrons (the kinetic freeze-out  problem
 \cite{KABkinFO1,KABkinFO2}). 
The existing versions of {the} HRGM  do not face {such a} problem, since {this model} has the hard-core repulsion only, while the mean-field interaction in it is set to zero \cite{Andronic06,Horn,KABOliinychenko:12,Bugaev13,Stachel:2013zma,Veta14,Bugaev:ALICE,VetaNEW,Bugaev17b}. Due to such a choice of interaction the HRGM  has the same energy per particle as an ideal gas and,  hence, it can be tuned to describe the  existing experimental hadronic multiplicities in central nuclear collisions from the lower AGS collision energy $\sqrt{s_{NN}} = 2.76$ GeV to the ALICE  center of mass energy $\sqrt{s_{NN}} = 2.76$ TeV with the total 
quality of fit  $\chi^2/dof \simeq 1.04$ \cite{Bugaev:ALICE,VetaNEW}.

Therefore,
any realistic hadronic EoS  of hadronic matter should be able to reproduce the pressure, entropy and all charge densities obtained by the  HRGM at the   chemical freeze-out curve  $\mu=\mu_{CFO}(T)$. In particular, for the  mean-field models discussed here it means that they should be extended in order to include all other hadrons and that  at the curve $\mu=\mu_{CFO}(T)$ the total interaction pressure must vanish, i.e.  $P_{int} =0$, since it does not exist in the HRGM.  

In other words, if  at  the chemical freeze-out   curve such a model EoS  has a  non-vanishing attraction, then  it must have an additional repulsion to provide $P_{int} =0$.
Only this condition will  help one  to avoid  a hard mathematical problem of kinetic freeze-out to convert  the interacting particles into
a gas of free streaming particles  \cite{KABkinFO1,KABkinFO2}, since the HRGM with the hard-core repulsion and with vanishing mean-field interaction has the same energy per particle
as an ideal gas. Due to its importance, in Appendix we  analyzed the IST EoS and show that this EoS  also posses such a property.
Also the condition $P_{int} =0$  at chemical freeze-out curve will  make  a direct  connection between the realistic EoS and the hadron multiplicities  measured in heavy ion collisions. It is clear, that without $T$-dependent mean-field interaction $P_{int}$ such a
condition  cannot be fulfilled.

{
Despite many valuable  results  obtained  with  the HRGM the hard-core radii are presently  well established for the most abundant hadrons only, namely for pions
($R_\pi \simeq 0.15$ fm), the lightest  K$^\pm$-mesons ($R_K \simeq 0.395$ fm),
nucleons ($R_p \simeq 0.365$ fm) and  the lightest (anti)$\Lambda$-hyperons ($R_\Lambda \simeq 0.085$ fm) \cite{Bugaev:ALICE,VetaNEW}.
Nevertheless, we  hope  that the new data of high quality  on the yields of many strange hadrons recently measured by the ALICE  Collaboration at CERN  \cite{ALICE17} at the center of mass energy $\sqrt{s_{NN}} = 2.76$ TeV and the ones which are expected to be measured
during the Beam Energy Scan II at RHIC BNL (Brookhaven) \cite{RHIC17}, and at the accelerators of new generations, i.e. at  NICA  JINR (Dubna) \cite{NICA,NICA2} and FAIR GSI (Darmstadt) \cite{FAIR,FAIR2} will help us to determine their hard-core radii with high accuracy.
}
We have to add only that the IST EoS for quantum gases is well suited for such a task due to additive pressure $p_{id}(T, \nu_{1, 2})$, whereas  the generalization of the CS EoS of Ref.  \cite{Vovch17} to a multicomponent case looks rather problematic, since the CS EoS \cite{CSEoS} is the one component EoS by construction.

\section{Conclusions}  The self-consistent generalization of the IST EoS for quantum gases is worked out. It is shown that with this EoS one can  go beyond the VdW approximation at any temperature.  The restrictions on the temperature dependence of the mean-field potentials are discussed. It is found  that at low temperatures these potentials  either should be $T$-independent or should vanish faster than the first power of temperature providing the fulfillment of the Third Law of thermodynamics.  The same is true for the quantum VdW EoS. Hence, the idea to improve the quantum  VdW EoS by tuning the interaction  part of  pressure \cite{Vovch14,Vovch15} is disproved for low temperature $T$: if this part of pressure is linear in  $T$, then the VdW EoS breaks down the Third Law of thermodynamics; if it vanishes faster than the first power of  $T$, then the interaction part of pressure is useless, since it vanishes faster than the first term of the quantum virial expansion.  An alternative EoS \cite{Vovch17} allowing one to  abandon  the VdW approximation for nuclear matter is analyzed here and it is shown that for the same parameters at low temperatures  the IST EoS is softer at packing fractions $\eta \ge 0.05$.

The virial expansions for quantum VdW and IST EoS are established  and  the explicit expressions for all quantum virial coefficients, exact  for VdW and approximative ones for {the} IST EoS,   are given.  Therefore, for the first time the analytical expressions for the third and fourth quantum virial coefficients
are derived  for  the EoS
which is more realistic than the VdW one.
The source of softness of the IST EoS is demonstrated using the effective virial expansion for the effective proper volume which turns out to be
compressible.
The generalization of the traditional virial expansions  for  the mixtures of particles with different hard-core radii is straightforward.

The general constraints on realistic EoS for nuclear and hadronic matter are discussed.
We hope that using
 the revealed  properties of  the IST EoS for quantum gases   it will  be possible to go far beyond the traditional VdW approximation
 and that  due to its advantages this EoS   will become a useful tool for
heavy ion physics and for nuclear astrophysics.
Furthermore, we hope that  the developed EoS will help us  to determine the hard-core radii of hadrons from the new high quality
ALICE data and the ones which will be measured at RHIC,  NICA and FAIR.
\\

{\bf Acknowledgments.}
The authors appreciate the valuable  comments of  D. B. Blaschke, R. Emaus, B. E. Grinyuk,  D. R. Oliinychenko  and D. H. Rischke.
K.A.B., A.I.I., V.V.S. and G.M.Z.   acknowledge  a partial  support of  the National Academy of Sciences  of Ukraine (project No. 0118U003197).
V.V.S. thanks the Funda\c c\~ao para a Ci\^encia e Tecnologia (FCT), Portugal, for the
partial financial support to the Multidisciplinary Center for Astrophysics (CENTRA),
Instituto Superior T\'ecnico, Universidade de Lisboa, through the Grant No.
UID/FIS/00099/2013.  The work of A.I. was performed within the project SA083P17 of Universidad de Salamanca
launched by the Regional Government of Castilla y Leon and the European Regional Development
Fund.

\section{Appendix}

Here we consider a special choice of the  mean-field potentials which are temperature independent, i.e. $U_A = U_A (n_{id\, A})$ and show that at low particle densities
the energy per particle of such an EoS coincides with the one of the ideal gas. The analysis is made for a single sort of particles, but it is evident  that generalization to the multicomponent case is straightforward.

For the considered  choice of  the  mean-field potentials  Eq.  (\ref{EqXXXVI}) for the entropy per particle becomes
\begin{eqnarray}
 \label{EqLX}
\frac{s_1} {n_1} &=& \frac{
\left[ \frac{s_{id\,1} } {n_{id\,1}} - 3 \, V_0  \, n_2 \cdot \frac{s_{id\,2}} {n_{id\,2}} \right]
}{ \left[1 -  3 \, V_0  \, n_2 \right]  } \simeq  \frac{s_{id\,1} } {n_{id\,1}} \,,
%
%%%
\end{eqnarray}
where in the first step we applied the relation $\tilde s_{id\,A} = s_{id\,A}$ with $A \in \{1; 2\}$ to Eq. (\ref{EqXXXVI}),
while  in the second step we used an approximation $\frac{s_{id\,2}} {n_{id\,2}} \simeq \frac{s_{id\,1} } {n_{id\,1}}$. The latter result follows from the condition  (\ref{EqXXIX}). Then in  the low density  limit, i.e. for  $e^\frac{\nu_2-\nu_1} {T} \simeq 1$, one gets
the relation  (\ref{EqXXVIII})  for the distribution functions $\phi_{id} (k, T, \nu_2) $ and  $\phi_{id} (k, T, \nu_1)$, which can be
approximated further on as $\phi_{id} (k, T, \nu_2) \simeq \phi_{id} (k, T, \nu_1)$  and, therefore,  one finds $p_{id}(T, \nu_2 ) \simeq   p_{id}(T, \nu_1  )$,  $n_{id}(T, \nu_2 ) \simeq   n_{id}(T, \nu_1  )$ and $s_{id}(T, \nu_2 ) \simeq   s_{id}(T, \nu_1  )$.

The energy per particle for the EoS  (\ref{EqXXII}) can be found from the thermodynamic identity
\begin{eqnarray}
\label{EqLXI}
&& \frac{\epsilon_1}{n_1} = T \frac{s_1}{n_1} + \mu - \frac{ p (T, \mu )  }{n_1} \,. ~
\end{eqnarray}
Expressing the chemical potential $\mu$ via an effective one $\nu_1$ from Eq. (\ref{EqXXIV})  one can write
$\mu = \nu_1 +  V_0 p_{id\,1}- V_0 P_{int\,1}  + 3V_0 p_{id\,2} - 3V_0 P_{int\,2}  - U_1$. Substituting this result into Eq. (\ref{EqLXI}), one finds
\begin{eqnarray}
\label{EqLXII}
\frac{\epsilon_1}{n_1} &\simeq& T \frac{s_{id\,1}}{n_{id\,1}} + \nu_1 - U_1 +  \left[  V_0 - \frac{ 1 }{n_1} \right] (p_{id\,1} -  P_{int\,1})  \nonumber ~\\
&+&  3V_0 (p_{id\,2} -  P_{int\,2}) \,,
\end{eqnarray}
where Eq.  (\ref{EqLX}) was  also used.  Approximating the particle number density $n_1$  in Eq.  (\ref{EqXXXV}) as
\begin{eqnarray}
\label{EqLXIII}
n_1 \simeq \frac{n_{id\,1}}{1 + V_0  \, n_{id\,1} + 3 \, V_0  \, n_2 } \,,
\end{eqnarray}
and substituting it into Eq. (\ref{EqLXII}), one obtains
\begin{eqnarray}
\label{EqLXIV}
\frac{\epsilon_1}{n_1} &  \simeq  &\frac{\epsilon_{id\,1}}{n_{id\,1}} + 3V_0 n_2 \left[ \frac{p_{id\,2}}{n_2} - \frac{p_{id\,1}}{n_{id\,1}} \right]  - U_1  \nonumber ~\\
&  - &\left[  V_0  - \frac{ 1 }{n_1} \right]  P_{int\,1} -  3V_0   P_{int\,2} \,,
\end{eqnarray}
where we applied the thermodynamic identity  (\ref{EqLXI}) to the energy per particle for the gas of point-like particles with the chemical potential $\nu_1$. To simplify the evaluation  for the moment we assume that all mean-field interaction terms obey  the following equality
\begin{eqnarray}
\label{EqLXV}
 \hspace*{-1.1mm} \frac{(1 -   V_0 n_1) }{n_1}   P_{int\,1} (n_{id\,1}) -  3V_0  P_{int\,2} (n_{id\,2})  = U_1  (n_{id\,1}) .~
\end{eqnarray}
Using in Eq.  (\ref{EqLXIV}) the first two terms of virial expansion (\ref{EqVIII}) for the pressures $p_{id\,1}$ and $p_{id\,2}$  and
Eq.  (\ref{EqXXXVa})  for $n_2$ one finds
\begin{eqnarray}
\label{EqLXVI}
&&\frac{p_{id\,2}}{n_2} - \frac{p_{id\,1}}{n_{id\,1}} \simeq T \left[ (1+ a_2^{(0)} n_{id\,2} )(1 + 3\alpha V_0 n_{id\,2}) \right. \nonumber \\
&&- \left. (1+ a_2^{(0)} n_{id\,1} ) \right] \simeq  T (1+ a_2^{(0)} n_{id\,1} )3\alpha V_0 n_{id\,1}\,,
\end{eqnarray}
where in the last step of derivation we used the low density approximation $n_{id\,2} \simeq  n_{id\,1}$. Finally, under the condition (\ref{EqLXV}) Eq. (\ref{EqLXIV}) acquires the form
\begin{eqnarray}
\label{EqLXVII}
\frac{\epsilon_1}{n_1} &\simeq& \frac{\epsilon_{id\,1}}{n_{id\,1}} + 9\alpha V_0^2 n_2 n_{id\,1} \, T \, (1+ a_2^{(0)} n_{id\,1} ) \,.
\end{eqnarray}
Since the typical packing fractions $\eta = V_0 n_1 \simeq V_0 n_2 \simeq V_0 n_{id\,1}$ of the hadron resonance gas model at chemical freeze-out do not exceed  the value 0.05 \cite{Bugaev:ALICE}, then the second term on the right hand side of Eq.  (\ref{EqLXVII}) is not larger than
\begin{eqnarray}
\label{EqLXVIII}
0.025 \alpha \,  T (1+ a_2^{(0)} n_{id\,1} ) \,.
\end{eqnarray}
Comparing this estimate with the energy per particle for the lightest hadrons, i.e. for pions, in the non-relativistic  limit
$\frac{\epsilon_{id\,1}}{n_{id\,1}}\bigl|_\pi \simeq m_\pi + \frac{3}{2}T$ (here $m_\pi \simeq 140$ MeV), one can be sure  that for temperatures at which the hadron gas exists, i.e. for $T < 160$ MeV,  the term  (\ref{EqLXVIII})   is negligible and, hence,  with high accuracy one finds $\frac{\epsilon_1}{n_1} \simeq\frac{\epsilon_{id\,1}}{n_{id\,1}}$.

Now let's discuss the condition  (\ref{EqLXV}).  It is apparent that in the general case it can hold, if the mean-field  interaction is absent, i.e. $U_1 = U_2 =0$ and $P_{int\,1} = P_{int\,2} =0$. This is exactly the case of the hadron resonance gas model.  However,  one might think that
there exist a special case for which Eq.  (\ref{EqLXV})  is the simple differential equation for two independent variables $n_{id\,1}$ and $n_{id\,2}$.  Let's show that this is impossible.
First, with the help of Eq.   (\ref{EqXXXV})
 we rewrite  the term $\frac{(1 -   V_0 n_1) }{n_1} =  [n_{id\,1} (1-3 V_0 n_2)]^{-1}$. Then Eq. (\ref{EqLXV}) can be cast as
\begin{eqnarray}
\label{EqLXIX}
\frac{P_{int\,1} (n_{id\,1})/ n_{id\,1}}{ (1-3 V_0 n_2 (n_{id\,2}))} -  3V_0  P_{int\,2} (n_{id\,2})    =  U_1  (n_{id\,1}) \,.
\end{eqnarray}
From this equation one sees that the only possibility to decouple  the dependencies on $n_{id\,1}$ and $n_2$ in the first term above is to assume that  $P_{int\,1} = C n_{id\,1}$  where $C$ is some constant.  However, in this case one finds that the $n_{id\,1}$-dependence
of the right hand side of Eq.  (\ref{EqLXIX}) remains, since $U_1 = C \ln (n_{id\,1})$. Therefore, there is a single possibility
to decouple  the functional dependence of  $n_{id\,1}$ from $n_2$, namely  that $C = 0$ which means that  $P_{int\,2} =0$.

One can, however,  consider Eq.  (\ref{EqLXIX}) under the low density approximation assuming that $n_{id\,2} =n_{id\,1}$.   In this case
Eq. (\ref{EqLXIX}) defines the functional dependence of  $P_{int\,2} (n_{id\,1})$ for any reasonable choice of  the potential $U_1(n_{id\,1})$. Note that in this case the function  $P_{int\,2} (n_{id\,1})$ can be rather complicated even for the simplest choice of  $U_1(n_{id\,1})$ and,
hence,  the practical realization of the dependence   (\ref{EqLXIX}) seems to be  problematic.  Therefore, the most direct way to avoid the problem to convert the interacting particles into the free streaming ones  \cite{KABkinFO1,KABkinFO2}, is to  use only the hard-core repulsion between hadrons and
set to zero all other  interactions  at chemical freeze-out.

%%%%%%%%%%%%%%%


\begin{thebibliography} {99}


\bibitem{GenRevEoS}
{N. K.  Glendenning},
{``Compact Stars'', {\rm Springer-Verlag, New York}} {(2000)}.


 \bibitem{Typel10}
%
{S. Typel, G. R\"opke, T. Kl\"ahn, D. Blaschke and H. H. Wolter},
 %%``Composition and thermodynamics of nuclear matter with light clusters,''
{Phys. Rev. C} {\bf 81}, {(2010)} {015803} and references therein.

\bibitem{David15}
%
{S. Benic, D. Blaschke, D. E. Alvarez-Castillo, T. Fischer and S. Typel},
{Astron. Astrophys.} {\bf 577}, {(2015)} {A40}.
 %%% S. Benic, D. Blaschke, D. E. Alvarez-Castillo, T. Fischer and S. Typel, Astron. Astrophys. 577, A40 (2015).

\bibitem{Walecka74}
%
{J. D. Walecka},
{Annals Phys.} {\bf 83},  {(1974)} {491}.

\bibitem{Zimany88}
%
{J. Zimanyi et al.},
{Nucl. Phys. A} {\bf 484}, {(1988)} {647}.

\bibitem{Bugaev89}
%
{K. A. Bugaev and M. I. Gorenstein},
{Z. Phys. C} {\bf 43}, {(1989)} {261}.

\bibitem{Rischke91}
%
{D. H.  Rischke,  M. I.  Gorenstein,  H.  St\"ocker and W.  Greiner},
{Z. Phys. C} {\bf 51}, {(1991)} {485}.

\bibitem{SkyrmeI}
%
T. H. R. Skyrme, Phil. Mag. 1, 1043 (1956); Nucl. Phys. 9, 615 (1959).

\bibitem{SimpleLiquids1}
%
J. P. Hansen and I. R. McDonald, {\it  ``Theory of  simple liquids"}, Academic, London (2006).

\bibitem{VanHoveI}
%
{L. Van Hove},
{Physica} {\bf 15}, {(1949)} {951}.

\bibitem{VanHoveII}
%
{L. Van Hove},
{Physica} {\bf 16}, {(1950)} {137}.

\bibitem{Rischke93}
%
{M. I. Gorenstein, D. H. Rischke, H. St\"ocker, W. Greiner and K. A.Bugaev},
{J. Phys. G} {\bf 19}, {(1993)} {L69}.

\bibitem{Vovch14}
%
{D. Anchishkin and V. Vovchenko},
{\rm arXiv:1411.1444 [nucl-th]}   and references therein.
%%%

\bibitem{Vovch15}
%
{V. Vovchenko, D. V. Anchishkin and M. I. Gorenstein},
{Phys. Rev. C} {\bf 91}, {(2015)} {064314}.

\bibitem{Redlich16}
%
{K. Redlich and  K. Zalewski},
{Acta Phys. Polon. B} {\bf 47}, {(2016)} {1943}.

\bibitem{CSEoS}
%
N. F. Carnahan and  K. E. Starling,  J. Chem. Phys. {\bf  51},  (1969) 635.


\bibitem{Huang}
%
{K. Huang},
{\it ``Statistical mechanics'', {\rm Wiley \& Sons}}(1963).

\bibitem{Andronic06}
%
{A. Andronic, P. Braun-Munzinger  and  J. Stachel},
{Nucl. Phys. A} {\bf 772}, {(2006)} {167} and references therein.

  \bibitem{Horn}
 %
K. A. Bugaev, D. R. Oliinychenko,   A. S. Sorin and G. M. Zinovjev,
%Simple Solution to the Strangeness Horn Description Puzzle,
Eur. Phys. J. A {\bf 49}, (2013) 30.

\bibitem{KABOliinychenko:12}
%
D. R. Oliinychenko, K. A. Bugaev and  A. S. Sorin,
%%%INVESTIGATION OF HADRON MULTIPLICITIES AND
%%%HADRON YIELD RATIOS IN HEAVY ION COllisions
Ukr. J. Phys.  {\bf 58},   (2013) 211.
%%%[arXiv:1204.0103 [hep-ph]].
%%CITATION = ARXIV:1204.0103;%%

\bibitem{Bugaev13}
%
{K. A. Bugaev et  al.},
{ Europhys. Lett.} {\bf 104}, {(2013)} {22002}.
%%%and references therein.

\bibitem{Stachel:2013zma}
  J.~Stachel, A.~Andronic, P.~Braun-Munzinger and K.~Redlich,
  %``Confronting LHC data with the statistical hadronization model,''
  J.\ Phys.\ Conf.\ Ser.\  {\bf 509},  (2014) 012019 and references therein.
 %%% doi:10.1088/1742-6596/509/1/012019
  %%[arXiv:1311.4662 [nucl-th]].
  %%CITATION = doi:10.1088/1742-6596/509/1/012019;%%

  \bibitem{Veta14}
 V. V. Sagun, Ukr. J. Phys. {\bf 59},  (2014) 755.

\bibitem{Bugaev:ALICE}
%
{K. A. Bugaev et  al.},
%%, V. V. Sagun, A. I. Ivanytskyi, I. P. Yakimenko, E. G. Nikonov, A.V. Taranenko  and G. M. Zinovjev}
Nucl. Phys.  A {\bf 970}, (2018) 133.
%%CITATION = ARXIV:1611.07349;%%

 \bibitem{VetaNEW}
 %
{V. V. Sagun et al.}, 
Eur. Phys. J. A {\bf 54}, (2018)  100.
%%%arXiv:1703.00049  [hep-ph].
%%CITATION = ARXIV:1703.00049;%%

\bibitem{Bugaev17b}
%
K. A. Bugaev et al.,  
 Phys. Part. Nucl. Lett. {\bf 15}, (2018) 210.
%%%arXiv:1709.05419 [hep-ph] (accepted to Phys. Part.  Nucl. Lett.).
%%CITATION = ARXIV:1709.05419;%%

\bibitem{Bugaev:13NPA}
%
{V. V. Sagun, K. A. Bugaev, A. I. Ivanytskyi and  I. N. Mishustin},
{Nucl. Phys. A} {\bf 924}, {(2014)} {24}.

\bibitem{Bugaev:SMMI}
%
K. A. Bugaev, M. I. Gorenstein, I. N. Mishustin  and  W. Greiner,
%%%Exactly Soluble Model for Nuclear Liquid-Gas Phase Transition,
Phys. Rev. C{\bf 62}, (2000) 044320.
%%CITATION = NUCL-TH 0005036;%%


\bibitem{SMM0}
%
J. P. Bondorf et al., Phys. Rep. {\bf 257},  (1995) 131.

\bibitem{Kostyuk00}
%
{A. Kostyuk, M. I. Gorenstein,  H. St\"ocker and W. Greiner},
{Phys. Rev. C} {\bf 63}, {(2001)} {044901}.

 \bibitem{HardD}
 %
J. Kolafa and M. Rottner,  Mol. Phys. {\bf 104},  (2006) 3435.
 
 \bibitem{HardS}
  B. Barboy and W. M. Gelbart, J. Chern. Phys.  {\bf 71},  (1979) 3053.

 \bibitem{Typel16}
S. Typel, Eur. Phys. J. A {\bf 52},  (2016) 16.

\bibitem{Aleksei17}
 %
 A. I. Ivanytskyi, K. A. Bugaev, V. V. Sagun, L. V. Bravina and E. E. Zabrodin, 
 %%%Influence of Flow Constraint on the Properties of Nuclear Matter Critical Endpoint, 
 Phys. Rev. C {\bf  97}, (2018) 064905. 
%%%{A. I. Ivanytskyi et al.}, arXiv:1710.08218 [nucl-th].
%%CITATION = ARXIV:17010.08218;%%

\bibitem{RelVDW}
%
K. A. Bugaev,
%%%The Van-der-Waals Gas EOS for the Lorentz Contracted Spheres,
{Nucl. Phys. A} {\bf 807},   (2008)  251 and references therein.
%%CITATION = NUPHA,A807,251;%%



\bibitem{Vovch17}
%
{V. Vovchenko}
 %``Equations of state for real gases on the nuclear scale,''
  Phys.\ Rev.\ C {\bf 96},   (2017)  015206.
%%%{\rm  arXiv:1701.06524 [nucl-th]}.

\bibitem{Pawel}
%
{P. Danielewicz, R. Lacey  and W. G. Lynch}
{Science} {\bf 298}, {(2002)} {1592}.



\bibitem{Dutra14}
M. Dutra et al.,
 Phys. Rev. C {\bf 90},   (2014)   055203 and references therein.


\bibitem{Menezes17}
 %
 O. Lourenco,  M. Dutra and D. P. Menezes,
%``Critical parameters of consistent relativistic mean-field models,''
  Phys.\ Rev.\ C {\bf 95}, (2017)  065212.
%%  doi:10.1103/PhysRevC.95.065212
%%  [arXiv:1704.05498 [nucl-th]].
  %%CITATION = doi:10.1103/PhysRevC.95.065212;%%

\bibitem{NS17}
%
V. V. Sagun and  I. Lopes,
Astrophys. J {\bf  850}, (2017) 75.
%%%Neutron stars: A novel equation of state with induced surface tension,
%%% arXiv: 1709.07898 [astro-ph] (2017) 6 p. (Accepted for APJ publication).


\bibitem{Raju}
%
{R. Venugopalan and  M. Prakash}
{Nucl. Phys. A} {\bf 546}, {(1992)} {718}.

\bibitem{Satarov10}
%
L. M. Satarov,  M. N. Dmitriev and I. N. Mishustin,
Phys. Atom. Nucl.  {\bf 72}, (2009)   1390.


\bibitem{KABkinFO1}
%
K. A. Bugaev,
%%%Shock-like Freeze-out in Relativistic Hydrodynamics,
Nucl. Phys. A {\bf 606}, (1996) 559.
%%CITATION = NUCL-TH 9906047;%%

\bibitem{KABkinFO2}
%
K. A. Bugaev,
%%%Relativistic Kinetic Equations for Finite Domains and Freeze-out Problem,
Phys. Rev. Lett. {\bf 90},  (2003) 252301 and references therein.
%%CITATION = NUCL-TH 0210087;%%

\bibitem{Satarov15}
  L.~M.~Satarov, K.~A.~Bugaev and I.~N.~Mishustin,
%%% Equation of state and sound velocity of a hadronic gas with a hard-core interaction,
  Phys.\ Rev.\ C {\bf 91} (2015),  055203.
  %%CITATION = ARXIV:1411.0959;%%

\bibitem{ALICE17}
%
N. Agrawal et. al. [ALICE Collaboration],
%%%Probing the hadronic phase with resonances of different life- times in Pb-Pb collisions with ALICE
arXiv:1711.02408v1 [hep-ex].

\bibitem{RHIC17}
%
L. Adamczyk et. al. [STAR Collaboration],
%%%Probing Parton Dynamics of QCD Matter with ? and ? Production
Phys. Rev. C {\bf 93}, (2016) 021903.
%%%arXiv:1506.07605v1 [nucl-ex].

\bibitem{NICA}
%
D. Rischke,
%%Exploring strongly interacting matter at high densities - NICA White Paper,
{Eur. Phys. J. A} {\bf 52}, (2016) 267.
% arxiv:1512.08046v2.% [hep-ph].

\bibitem{NICA2}
%
P.  Senger,
%%%Nuclear  matter physics at  NICA,
{Eur. Phys. J. A} {\bf 52}, (2016) 217.

\bibitem{FAIR}
%
P.  Senger,
%%%The Compressed Baryon Matter experiment at FAIR,
{Nucl. Phys. A} {\bf 862-863}, (2011) 139.

\bibitem{FAIR2}
%
T. Ablyazimov  et al.,
%%%Challenges in QCD matter physics - The scientific programme of
%%%the Compressed Baryon Matter experiment at FAIR,
{Eur. Phys. J. A} {\bf 53}, (2017) 60.
% arxiv:1607.01487.% [nucl-ex].


\end{thebibliography}
\end{document}